\documentclass{aa}
\def \th {\thinspace}

\def \src {Sco\th X-1}
\def \degmark{^\circ}

\def \arcsec {\hbox{$^{\prime\prime}$}}

\def\approxgt{\mathrel{\hbox{\rlap{\lower.55ex \hbox {$\sim$}}
\kern-.3em \raise.4ex \hbox{$>$}}}}
\def\approxlt{\mathrel{\hbox{\rlap{\lower.55ex \hbox {$\sim$}}
\kern-.3em \raise.4ex \hbox{$<$}}}}
\def \th {\thinspace }
\def \ref {\reference{}}

\def \sun {\hbox {$\odot$}}
\def \degmark{^\circ}

\def \arcsec {\hbox{$^{\prime\prime}$}}
\usepackage{epsf}
\usepackage{graphicx}
\begin{document}

\thesaurus{6(13.25.5;  % X-rays: stars,
             08.09.2: % stars: individual: Sco\th X-1,
             08.14.1;  % stars: neutron,
             08.02.1;  % binaries: close,
             02.01.2)} % accretion, accretion disks

\title{Physical changes during Z-track movement in Sco\th X-1 on the flaring branch}

\author{ R. Barnard\inst{1}
\and M. J. Church\inst{2,3}
\and M. Ba\l uci\'nska-Church\inst{2,3}}

\offprints{R. Barnard@open.ac.uk}

\institute{Department of Physics and Astronomy, Open University,
           Milton Keynes, MK7 6AA, UK\\
\and
           School of Physics and Astronomy, University of Birmingham,
           Birmingham, B15 2TT, UK\\
\and
          Astronomical Observatory, Jagiellonian University, ul. Orla 171, 30-244 Cracow, Poland.\\}
                    
%\thanks{}

\date{Received 15 November 2002; Accepted 6 April 2003}
\authorrunning{Barnard et al.}
\titlerunning{Flaring in Sco\th X-1}
\maketitle

\begin{abstract}
We present results of a detailed study of X-ray flaring in the Z-track source Sco\th X-1
in a highly super-Eddington state
made using high quality {\it Rossi-XTE} data from the PCA and HEXTE instruments.
The emission model successfully used to explain the dipping LMXB, and other classes of LMXB
in recent years, was applied to study the physical evolution along the Z-track
which remains a major problem. This model consists of blackbody emission from the 
neutron star plus Comptonized emission from an extended accretion disk corona.
%The model provides good fits to all spectra, and no high energy tail was present
%as claimed by d'Amico et al. (2001) as an excess above a bremmstrahlung model fitted.
As found in earlier work, major changes take place in the neutron star blackbody 
emission with $kT$ increasing in flaring, and the blackbody radius $R_{\rm BB}$ increasing
substantially to a maximum value of 9.4$\pm$0.6 km, consistent with the radius of the neutron star,
after which $R_{\rm BB}$ decreases. Thus this result is a measurement of
neutron star radius. The behaviour of Sco\th X-1 in flaring is compared with our previous results for
the strong flaring that takes place in the bright dipping, flaring LMXB X\th 1624-490.
Remarkably, during movement along the Normal Branch towards the apex
with the Flaring Branch, the luminosities of both spectral components decrease,
suggesting the possibility that $\dot M$ may {\it decrease} on the
Normal Branch, contrary to the widely-held view that 
$\dot M$ increases monotonically along the Z-track. During flaring, we detect for the
first time an increase of the Comptonization cut-off energy which may suggest heating of the ADC
plasma by the neutron star flare. The energy of a broad Gaussian line at 
$\sim$6.4 keV does not change, but the intensity of the line increases in flaring
suggesting either an increase in ADC size in flaring or the effects of irradiation by the neutron 
star.
\end{abstract}


\section{Introduction}

Flaring in Sco\th X-1 has been known for many years, since it became apparent that it
had a quiescent state and an active or flaring state (Canizares et al. 1973; White et al. 1976).
In the former, variability was limited, whereas in the flaring state, strong intensity
variations are seen on timescale of minutes to hours. 
The similarity of other sources to Sco\th X-1 was realized by Mason et al. (1976),
and van Paradijs \& Lewin (1986) concluded that there was a distinction in 
Low Mass X-ray Binary sources (LMXB) between
low luminosity, bursting sources and the bright sources such as Sco\th X-1.

Study of sources like Sco\th X-1 revealed different patterns of correlated hardness-intensity
behaviour, leading to the conclusion of Hasinger et al. (1989) that the
bright sources can be found on three different branches forming a skewed Z-shape in
a hardness-intensity diagram: the Horizontal, Normal and Flaring branches. The Z-track sources 
are \hbox{Sco\th X-1,} \hbox{Cyg\th X-2}, GX\th 17+2, GX\th 349+2, GX\th 340+0 and \hbox{GX\th 5-1.}
%(Schulz, Hasinger \& Tr\" umper, 1989)
These hardness-intensity plots provided a model-independent method demonstrating that strong 
spectral changes were taking place within the sources, but not revealing the nature of the changes.
At the same time it was found that the pattern of quasi-periodic oscillations (QPO)
depended on position on the Z-track (van der Klis et al. 1987; Hasinger et al. 1989).
Although these changes are suggestive of changes in the neighbourhood of the inner disk
and compact object, they do not in themselves reveal the nature of the changes.
More recently, evidence has been found that some of the related Atoll sources
also move on a Z-track but on longer timescales (Muno et al. 2002); however,
van der Klis (2002) has argued that the pattern of timing behaviour on the ``Horizontal Branch''
in one atoll source 4U\th 1608-52 is not as expected for a Z-track source.
The Z-track and related Atoll sources have been seen generally as 
low-inclination LMXB sources
as distinct from the high-inclination dipping and ADC sources (Frank et al. 1987).
However, three of the Z-track sources display X-ray dipping implying
high inclination (e.g. Kuulkers et al. 1996).

The nature of the physical changes taking place along the Z-track is not known.
Detailed models have been proposed (e.g. Psaltis et al. 1995), but there has been no way
of testing the details of, for example, the geometry of such models against 
observation. However, it was suggested that
during movement along the  Z-track, a single parameter changes, and it 
has generally been assumed that
this is the mass accretion rate $\dot M$ as proposed by Hasinger et al. (1989).

Spectral modelling of Sco\th X-1 was carried out by White et al. (1986) using a
two-component model consisting of a blackbody plus a Comptonization term, and it was shown that
flaring consisted of increases of the blackbody luminosity. However, different
models were then proposed for LMXB which were radically different physically.
The Eastern model of Mitsuda et al. (1989) consisted of multi-colour blackbody emission from the 
accretion disk, plus Comptonized emission of the neutron star. The Western model of White 
and co-workers (e.g. White et al. 1986) was of a self-Comptonized or generalized 
thermal type to explain the dominance of 
Comptonization in LMXB emission, while bright sources required an additional blackbody
component (White et al. 1988). The ability of these different forms to fit
LMXB spectra has resulted in the nature of the emission remaining controversial to
the present.

\section{The Birmingham emission model}

In fact, the dipping LMXB sources displaying dips in X-ray intensity at the orbital period
due to absorption in the outer disk (White \& Swank 1982; Walter et al. 1982) have held the key 
to this problem. Spectral models are more strongly constrained by the requirement of having
to fit not only the non-dip spectrum, but also several intensity levels during dipping
without the source emission parameters changing. Work on the dipping sources led to a
model consisting of blackbody emission from the neutron star plus Comptonized emission
from an extended accretion disk corona (ADC) (Church \& Ba\l uci\'nska-Church 1995).
This model provides very good fits to all of the $\sim$10 dipping sources (e.g.
Church et al. 1997, 1998a, 1998b; Ba\l uci\'nska-Church et al. 1999, 2000; Smale et al.
2001, 2002; Barnard et al. 2001). It also
fits well the Atoll and Z-track sources included in an {\it ASCA} survey of LMXB
(Church \& Ba\l uci\'nska-Church 2001) showing that the model describes well all classes of LMXB
and that a blackbody component is present in all sources to different degrees. 

After an initial period of testing the Birmingham model with data from many observations of LMXB,
it was realized that use of the model revealed new features relating firstly to the blackbody
emission and secondly to the Comptonized emission of the ADC. Firstly, the {\it ASCA} survey
revealed systematic behaviour of the blackbody luminosity $L_{\rm BB}$. The factors governing
the level of observed thermal emission in LMXB have never been understood: Newtonian theory
suggests that 50\% of the X-ray emission should originate on the neutron star (or boundary layer).
Apart from bright sources, the level is generally much less than 50\%. The survey revealed
a new empirical relation that the height of the equatorial emitting strip on the neutron star $h$
was equal to the height of the inner, radiatively-supported accretion disk $H$ (Church \& 
Ba\l uci\'nska-Church 2001). We do not think that this approximate equality spanning more than
3 decades of luminosity is coincidental, but that it suggests two mechanisms, at least, that
actually determine the level of $L_{\rm BB}$. The first is the process of accretion flow 
spreading on the surface of the neutron star proposed by Inogamov \& Sunyaev (1999) which can 
explain the equality, since the extent of vertical
spreading depends on the mass accretion rate, so that $h$ depends on the luminosity, as does $H$.
A comparison of the {\it ASCA} survey results with this theory (Church et al. 2002) 
was encouraging, with good agreement at low luminosities and agreement within a factor of three
at higher luminosities. The other possibility is that material flows advectively, radially
from the inner edge of the disk to the neutron star, directly leading to an approximate equality
$h$ = $H$ as observed.

Our observations of the dipping LMXB have provided a significant result on the nature
of the Accretion Disk Corona. Spectral fitting shows that this region is {\it clearly} 
extended as this emission component is removed very gradually as dipping progresses, 
the level only falling to zero when 
the absorbing region completely overlaps the extended emission region at the centre of a dip
(Church et al. 1997). Furthermore,
dip ingress times allow direct measurement of the size of the extended emitter when the
absorber is known to have larger angular extent than the emitter, i.e. when the intensity falls
to zero in dipping in any energy band. By this technique, the radius of the ADC
has been measured in a number of dipping sources, and found to be typically $\sim$50,000 km
or 15\% of the accretion disk radius (Church 2001). The result is robust and precludes all
models with small, central Comptonizing ADC including the Eastern model, since this would have size
$\approxlt$100 km, i.e. 1000 times smaller. 

\section{The Birmingham model versus the Eastern model}

The controversy over the emission model for LMXB persists to the present, various
workers applying the Eastern model, while we have obtained strong evidence that the
Birmingham model is correct. The Birmingham model differs from the Western model
in that the evidence is for a blackbody component in {\it all} sources.
In the Eastern model, X-ray emission is assumed to consist of
disk blackbody emission plus neutron star emission Comptonized in an inner region.
We summarize the evidence for and against each model, which we then discuss
in more detail. One aspect of the Eastern model is that it often justified on the
grounds of theoretical expectations (e.g. Done et al. 2002),
not on observational evidence.

\vskip 1 mm \noindent
Evidence for the Eastern model:

\vskip - 3mm
\begin{enumerate}
\vskip - 3 mm
\item Its ability to fit LMXB spectra
\end{enumerate}

\vskip -1 mm\noindent 
Evidence against the Eastern model:
\vskip - 3mm
\begin{enumerate}
\vskip - 3mm

\item Dip ingress timing proves that the ADC is large

\item Smooth transition of blackbody properties from burst to non-burst emission supports blackbody 
origin on the neutron star

\item Unphysically small values of inner disk radii $<<$ radius of the neutron star when fitting
the Eastern model
\end{enumerate}

\vskip - 1mm\noindent
Evidence for the Birmingham model:
\vskip - 3mm
\begin{enumerate}
\vskip - 3mm
\item Its ability to fit LMXB spectra

\item The undoubted presence of extended Comptonized emission in the spectra of the dipping LMXB

\item Dip ingress timing {\it proves} the ADC is extended

\item The consistency of the derived blackbody radius with the radius of the neutron star 
in all sources
up to a maximum value of $\sim$10 km as in Sco\th X-1 (the present work)

\item Assuming blackbody origin on the neutron star in the {\it ASCA} survey leads to
reasonable explanations of the mechanism determining the blackbody luminosity for the first time
(Sect. 2).

\end{enumerate}

The measurement of dip ingress times is strong, direct evidence for an extended, flat
ADC above the accretion disk many times larger than required by the Eastern model.
The only way of avoiding this conclusion would be if the absorbing region in dipping sources
on the outer disk had a complex structure that caused gradual removal of point-like
ADC emission. However, attempts to model gradual dip ingress in the source XB\th 1916-053
on this basis showed that the observed dipping could not be reproduced (\.Zycki, private 
communication 2001). 

Secondly, our {\it ASCA} survey of LMXB included testing of the Eastern model. Results of spectral fitting
a two-component model were presented (Church \& Ba\l uci\'nska-Church 2001), this model consisting
of a cut-off power law representation of Comptonization plus disk blackbody. In many cases, 
the level of this thermal component was such that the radius of the inner disk $r_{\rm i}$,
which is a parameter of the disk blackbody, was less than 0.5 km, and so unphysical
as many times less than the neutron star radius of 10 km (Church 2001). The validity of
this conclusion was questioned by Done et al. (2002) in their paper based on fitting
the Eastern model to the Z-track source Cyg\th X-2. They argued that 
the very small $r_{\rm i}$ were an artifact of using an incorrect
Comptonization model, i.e. a cut-off power law. This was based on assuming that
the seed photons came from the neutron star or inner disk, and should be modelled
by a simple blackbody, having the effect of a low energy cut-off in the
Comptonized spectrum, which would fall steeply below 1 keV. Use of a cut-off power law
would overestimate the Comptonized emission, and the disk blackbody would
be underestimated. They concluded that the Eastern model was probably correct.
The argument is however, not correct since the spectrum of the Comptonized emission
depends on the {\it size of the ADC}. For an ADC extending 15\% or more across the disk, the
seed photons consist of emission from the inner 15\% of the disk, and so have a soft 
integrated spectrum. The Comptonized spectrum thus continues to rise with approximately power law form 
at energies below 1 keV to 0.1 or even lower (Church et al. 2002). 
Thus the argument of Done et al. (2002) is not correct, and the small values of $r_{\rm i}$ are
not an artifact, and such results remain inconsistent with the actual radius of the inner disk.
In the present work, we have applied both the Birmingham model and a two-component model
in which the Comptonization term has disk blackbody seed photons, and find that the results 
of the two models are very similar, demonstrating that use of a cut-off power law is valid 
if the ADC is extended (Sect. 5.4).

If we accept the extended nature of the ADC as proven, the difference between
the Eastern and Birmingham models reduces to whether the blackbody emission is from the
inner disk or the neutron star. To avoid the observed emission being from the neutron star,
the Eastern model postulates that this is Comptonized. Direct observational testing is
difficult. However, the reasonable agreement between our results and the theory of
Inogamov \& Sunyaev supports origin on the neutron star. The large measured sizes of
the ADC, and the high optical depth of the ADC (Church 2001) suggests that disk blackbody
emission {\it will not be seen} since all of the hot inner part of the disk is covered
by reprocessing corona. There are other arguments in favour of neutron star blackbody emission
which we will only mention here:
e.g. the blackbody X-ray emission in X-ray bursts decays to the quiescent blackbody component, 
and there is no doubt that the burst emission is from the surface of the neutron star.

Finally, an argument that has been raised against the Birmingham 
model is that theoretically the neutron star blackbody should be modified substantially by 
electron scattering in the atmosphere of the neutron star. We have previously addressed
this argument (Ba\l uci\'nska-Church et al. 2001), making the point that the only evidence for such modification
is in a fraction of X-ray bursts (having $kT$ $\sim$3 keV), there being no evidence in
non-burst emission, and that theoretically whether modification takes place depends on the
electron density which is very poorly known.

Thus there is substantial evidence in favour of the Birmingham model, and it is clearly of
interest to test whether this model can provide an explanation of Z-track behaviour, i.e. show 
what physical changes take place during track movement. It is also plain that this can only be done when
the correct model is used. Since the early work on Sco\th X-1, there has been relatively little 
spectral fitting work aimed at solving the Z-track problem. In the present paper, we apply the 
Birmingham model to X-ray spectra of Sco\th X-1 on the Normal and Flaring Branches using high quality 
data from {\it Rossi-XTE}.

\section{Observations and data analysis}

Numerous observations of Sco\th X-1 have been made using {\it RXTE}
(Bradt et al. 1993); we have selected an observation made
in two parts on 1998, January 7 and 8, since in the first observation, the
source was in a strongly flaring state, but in the second was not
flaring. Each observation lasted $\sim$35 ks. 
The data presented here were collected with both the proportional counter array (PCA) 
and the high energy X-ray timing experiment (HEXTE), using the PCA 
instrument in Standard 2 mode with a time resolution of 16~s. The PCA
consists of five Xe proportional counter units (PCU) numbered 0--4,
with a combined total effective area of about 6500 cm$^2$ (Jahoda et
al. 1996). Examination of the housekeeping data showed that
PCU 0, 1, and 2 were reliably on during both observations,
and PCU 3 and 4 were off. However, the count rate of the source Sco\th X-1
was so high that 
\begin{figure*}[!ht]                                             %Fig. 1
\begin{center}
\includegraphics[width=90mm,height=174mm,angle=270]{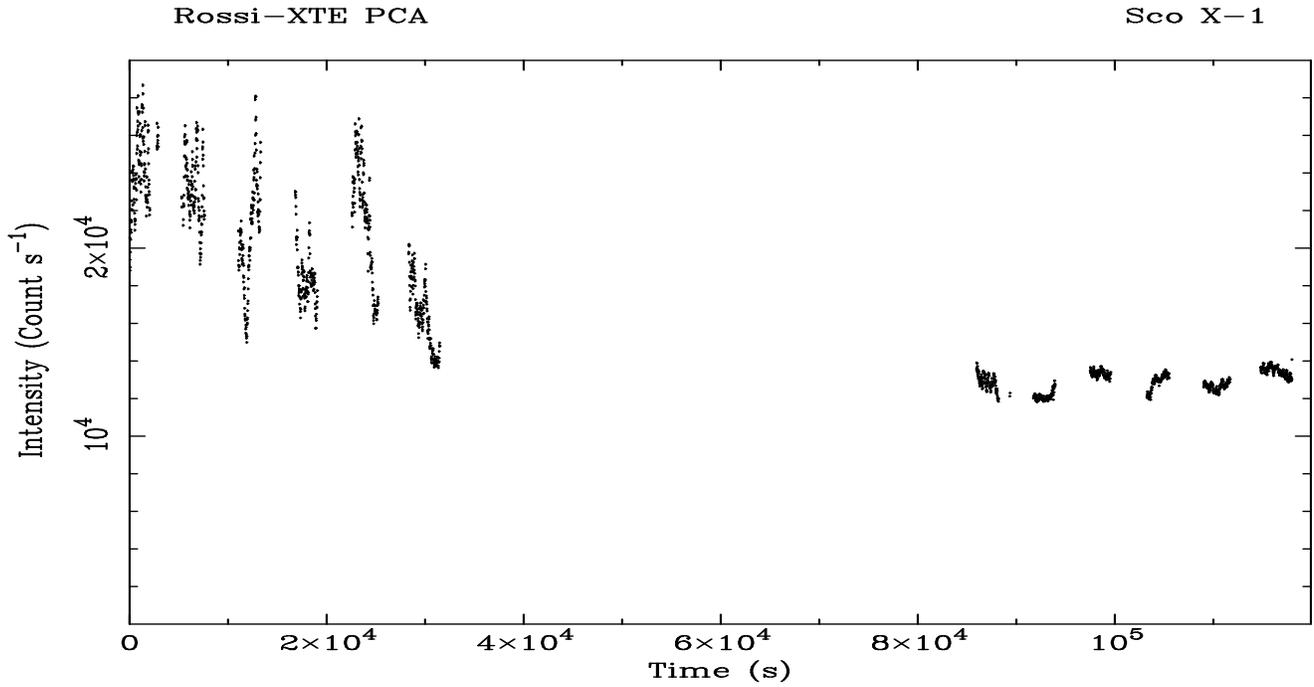}     %was tot
\caption{Background-subtracted, deadtime corrected PCA 1.9--18.5 keV light curve
of the 1998, Jan 7--8 {\it RXTE} observation of Sco\th X-1 with 16 s binning}
\end{center}
\end{figure*}
\begin{figure*}[!ht]                                             %Fig. 2
\begin{center}
\includegraphics[width=100mm,height=174mm,angle=270]{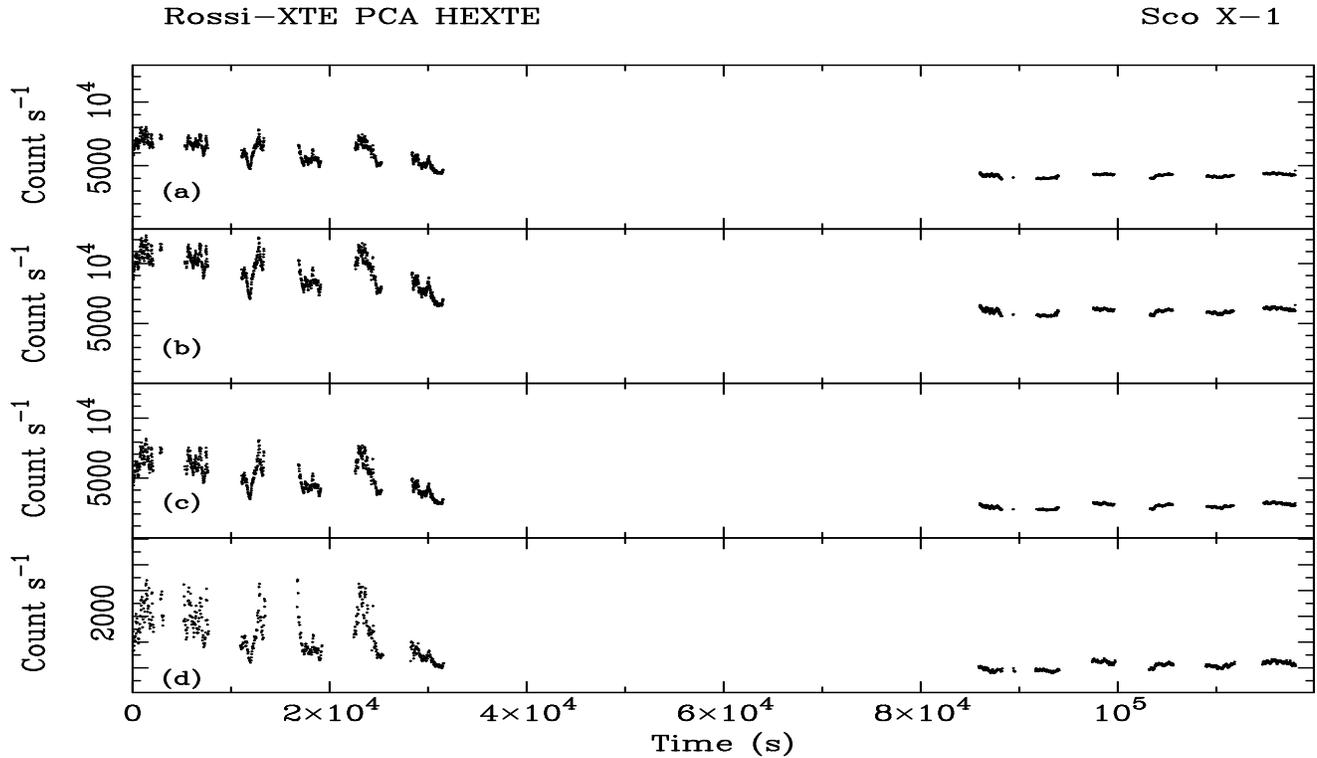}     % was allf
% allf.qdp = lbd.qdp, mbd.qdp, hbd.qdp + hextelc.qdp=tes1.qdp with odd points cleaned in 
% allf.qdp in window 4 = hexte LC
\caption{PCA light curves in the energy bands: (a) 1.9--4.1, (b) 4.1--7.3 and
(c) 7.3--18.5 keV plus (d) HEXTE light curve of the C1 cluster}
\end{center}
\end{figure*}
it was not necessary or desirable to use more than a single PCU, 
since this would complicate analysis, and so PCU0 data were chosen. These 
observations took place before the loss of the propane layer in PCU0 which 
affected the background count rate and channel-energy conversion for this detector.
Moreover, it was not necessary to use both anodes, and the left, but not the right, 
detector was used. Light curves and spectra were extracted as described below,
and background subtracted, and deadtime corrected. Spectra were also corrected
for pulse pileup necessary at the high count rate of Sco\th X-1.

Data were also obtained using the HEXTE phoswich detectors, which are
sensitive over the energy range 15--200~keV, and data from one of the two
clusters: Cluster 1, were selected.
%Each cluster consists of 4 detectors,
%but in Cluster 1, detector 2 had failed on March 6, 1996, before the present
%observations and so was excluded from the analysis.
The instrument was operated in rocking mode moving in a cycle: on-source,
offset by +1.5$\degmark$, on-source and offset by -1.5$\degmark$, spending
16 s in each position, so that the 
complete cycle occupied 64 s.
Deadtime-corrected HEXTE lightcurves and source and background spectra were obtained using the
{\sc hxtlcurv} tool with 32~s time binning.

\section{Results}

\subsection {The X-ray lightcurves}

The PCA light curves and spectra were extracted using the standard {\it RXTE} analysis 
software, {\sc ftools 5.2}. PCA lightcurves were extracted from the raw data
in Standard 2 mode 
for the top layer of the PCU0 instrument and the left anode. Standard screening was
applied to select only data having an offset of the telescope pointing axis from
the source less than 0.02$\degmark$, and elevation above the limb of the Earth
greater than 10$\degmark$. A total light curve in the band 1.9--18.5 keV was extracted, and
background subtraction carried out. The program {\sc pcabackest} was used to  
generate background data files corresponding to all PCA Standard 2 mode raw data,
which were then used for correction of light curves. The latest background model
was applied, specifically, the ``bright'' model recommended for Epoch 3 of the mission
(defined as 1996, April 15 -- 1999, March 22), produced in 2002.
%pca\_bkgd\_cmbrightvle\_e3v20020201.mdl.
The light curve was background subtracted, and then deadtime correction carried out
using dedicated software applying the prescription provided by the mission specialists.
The background-subtracted deadtime-corrected light curve is shown on Fig. 1 
This clearly shows strong flaring in the first part of the observation (Jan 7).
Individual large flares last several thousand seconds, however, more rapid
variability can also be seen in the light curve on timescales of 100--200 seconds.
There is also a systematic decrease in intensity which can be seen to continue
at the start of the Jan 8 data, after which intensity fluctuations become
relatively small compared with the earlier strong flaring.

Light curves were also extracted in a low, a medium and a high energy band,
suitably chosen to allow derivation of hardness-intensity diagrams.
The energy bands
\begin{figure}[!ht]                                             %Fig. 3
\begin{center}
\includegraphics[width=100mm,height=88mm,angle=270]{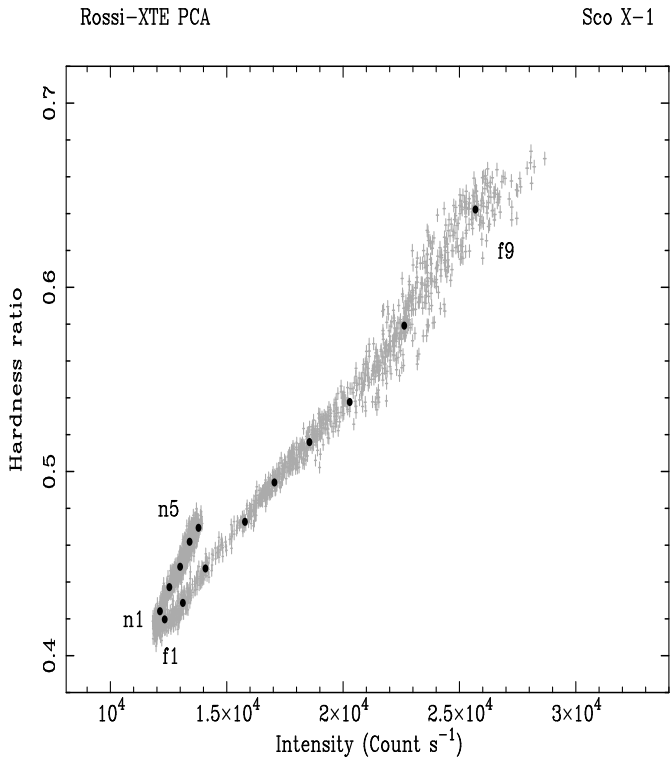}     % was comb
\caption{Hardness ratio as a function of intensity in the band 1.9--18.5 keV; the hardness
ratio was the ratio of the background-subtracted, deadtime-corrected
lightcurves in the bands 7.3--18.5 keV and 4.1--7.3 keV. Also shown superimposed
are points giving the mean hardness ratio and intensity of data selected in
intensity bands for spectral analysis labelled f1--f9 on the Flaring Branch 
and n1--n5 on the Normal Branch}
\end{center}
\end{figure}
used were 1.9--4.1 keV (low), 4.1--7.3 keV (medium) and 7.3--18.5 keV 
(high), which selecting the channel-to-energy conversion appropriate
to Epoch 3, correspond to the channel ranges 0--10, 11--19 and 20--50.
Fig. 2 (a)--(c) shows the background-subtracted, deadtime-corrected light curves in 
these bands.
A lightcurve was also extracted for standard-mode data from Cluster 1 of the HEXTE
instrument with at time resolution of 32 s using the program {\sc hxtlcurv};
this also carries out background subtraction and deadtime correction, and the latest
deadtime coefficients file released in February, 2000, applied.
The background subtracted, deadtime corrected lightcurve is shown in Fig. 2d in the total band,
effectively $\sim$18--50 keV in this source. 
Fig. 2 demonstrates dramatically
the high-energy nature of flaring. In the low energy band (Fig. 2 a), the intensity 
increases by 50\%, while in the high band (Fig. 2c), the increase is 100\% and in HEXTE 
(Fig. 2d) the increase is by 250\%. The spectra show that most of the HEXTE counts are 
in the band 20--50 keV.

\subsection{Hardness ratios}

Hardness ratios were formed by dividing the intensities in the 7.3--18.5 keV and 
4.1--7.3 keV bands, i.e. the high and medium bands. The variation of the hardness 
ratio with the summed intensity in the three bands spanning 1.9 to 18.5 keV is shown in Fig. 3
which also shows data selected for spectral analysis (below).
The hardness ratio clearly shows two well-defined tracks, corresponding to the
Normal and Flaring branches of the Z-track, containing the soft apex between these,
but without horizontal track data or the hard apex. Clearly, Sco\th X-1 underwent 
a transition between the Flaring Branch and the Normal Branch during the observation.
The location of the transition in the overall light curve (Fig. 1) was located
by examining each individual data section (separated by data gaps due to SAA passage 
and Earth occultation). For each section, a hardness-intensity plot was made and
overlayed on Fig. 3, allowing the section to be located on either the Flaring or Normal Branch.
This revealed that the transition took place at a time of 95 ksec, so that all data
before this was on the Flaring Branch, and all data later than this on the Normal Branch.
Before the transition, the source oscillates backwards and forwards
on the Flaring Branch. However, close inspection of the total light curve (Fig. 1)
shows that even on the Normal Branch, several flare-like variations of small amplitude continue, 
each lasting $\sim$1000 seconds.

\subsection{Selection of spectra}

Data were next chosen for spectral analysis so as to span each branch with
a number of spectra. Firstly, Flaring Branch data were selected by filtering
for times less than 95 ksec. Then 9 intensity bands were chosen and the corresponding
time intervals containing data in each band determined. These time filters
(GTI: Good Time Intervals) were then used to produce light curves and spectra 
(f1 -- f9) for each 
of the selections. Similarly, 5 bands were chosen for Normal Branch data (n1 --n5). 
Each selection was tested by overlaying the selected data as a mean value
on the hardness-intensity diagram of Fig. 3.
It can be seen that the selections used were satisfactory in that the points
lie along the centres of the tracks.
We have carried out a substantial amount of testing of this procedure, using the
present data, and data on other Z-track and atoll sources. As the objective is to
reveal the physical changes taking place during track movement, it is important
that spectral points lie approximately along the centre of tracks. If this was not the
case, spectral fitting results were found to change less systematically, as
motion perpendicular to the track as well as parallel was taking place. 

For each selection, spectra were extracted for the PCA using Standard 2 data,
and corresponding background spectra generated. 
\begin{figure}[!ht]                                             %Fig. 4
\begin{center}
\includegraphics[width=90mm,height=88mm,angle=270]{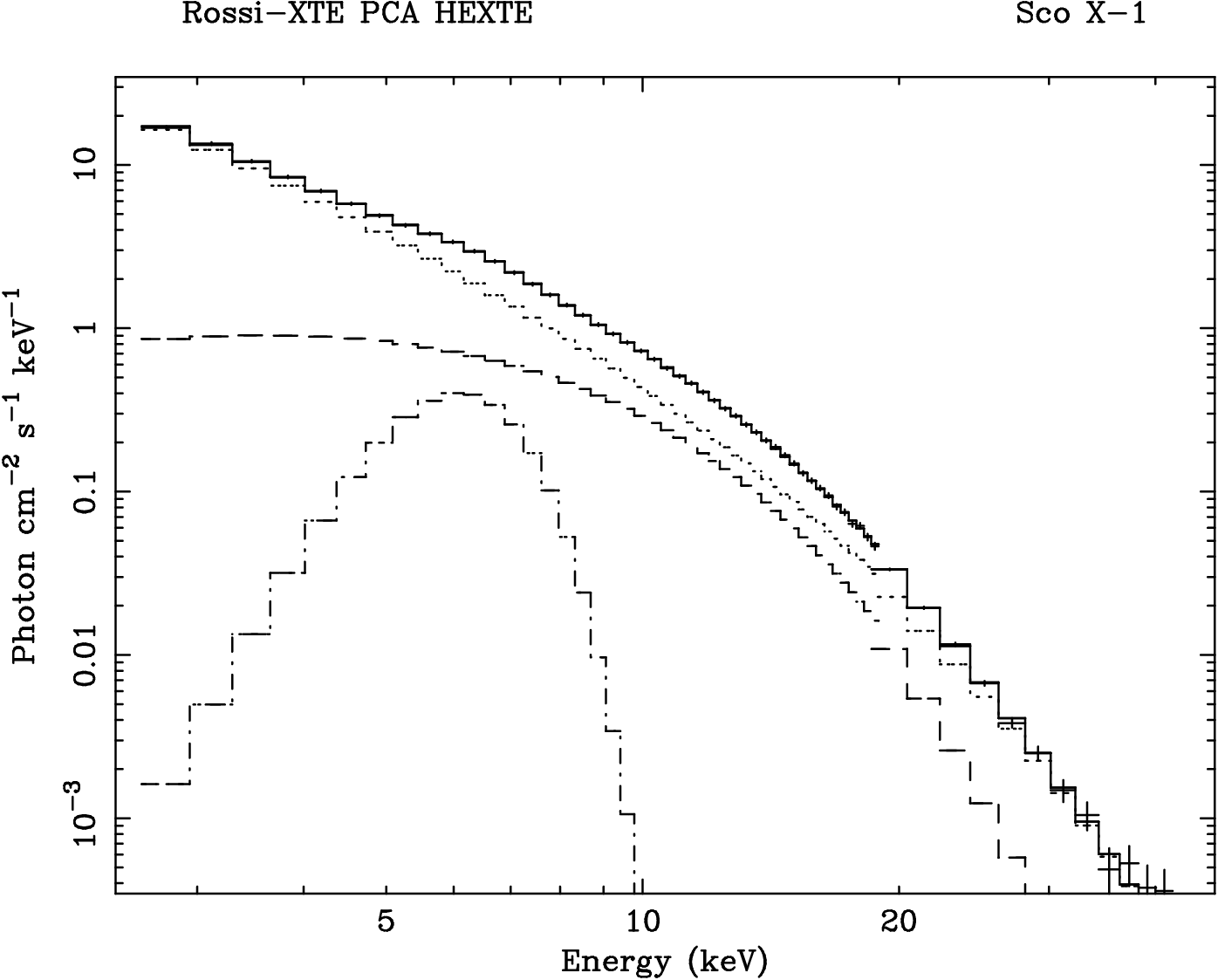}     %was pdatn0
\includegraphics[width=90mm,height=88mm,angle=270]{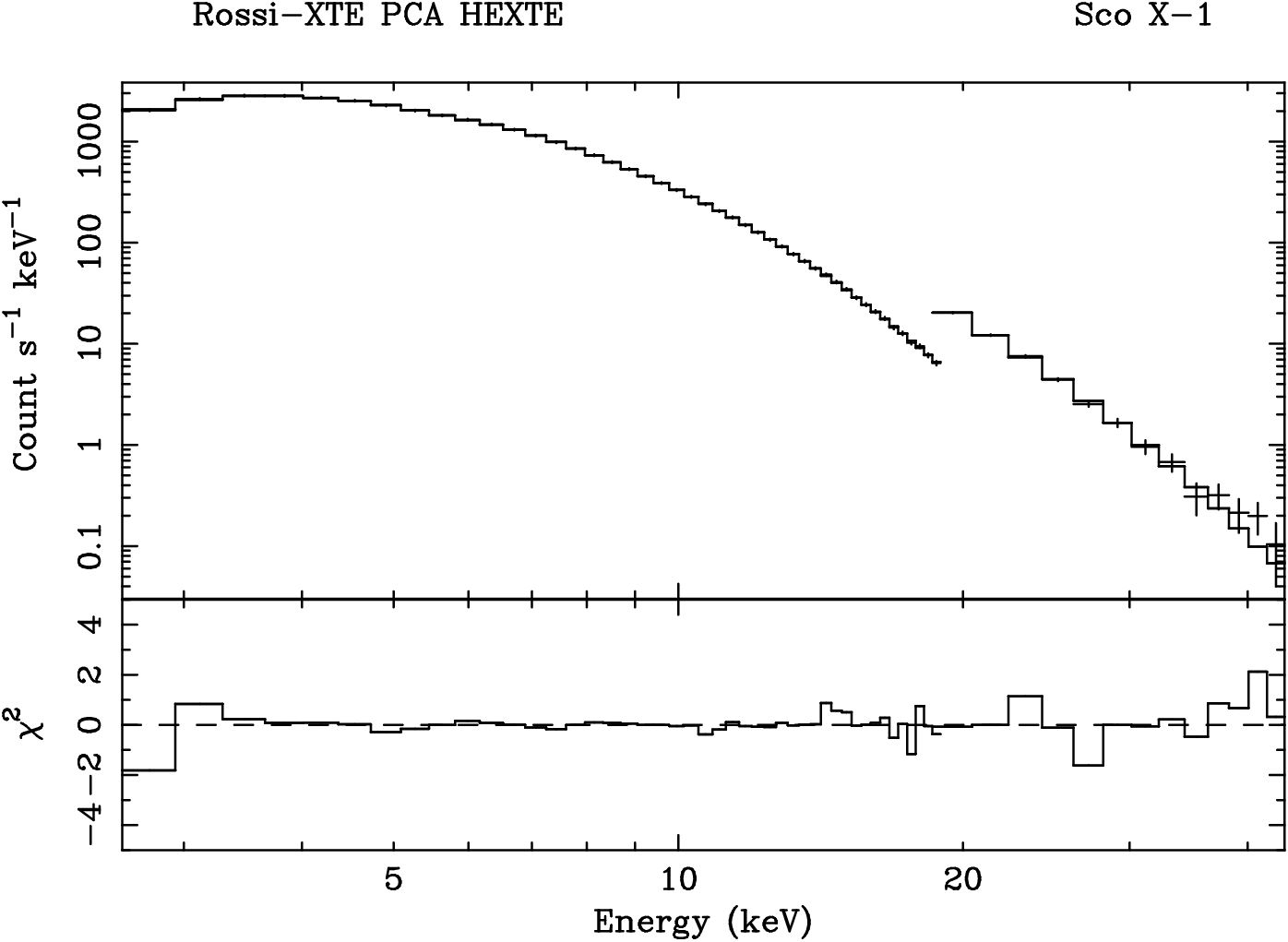}     %was ldatn0
\caption{Unfolded spectra (upper) and folded spectra (lower) for the best simultaneous fit to 
PCA and HEXTE data in a typical case: Normal Branch spectrum n1}
\end{center}
\end{figure}
Deadtime correction was performed,
and finally a correction made for pulse pileup effects using dedicated software, 
which is a significant correction in the case of Sco\th X-1. A time constant
of 1.80 $\mu$sec was used as the best value for high count rates
in the left anode detector (Jahoda et al. 1996). 
The high count rates
resulted in the Poisson errors in spectral channels being very small, and
so a
\begin{table*}
\caption{Spectral fitting results: blackbody temperature $kT$ and normalization, power law photon
index $\Gamma$ and normalization, line energy $E_{\rm l}$ and normalization. 90\% confidence
errors are shown}
\begin{center}
\begin{minipage}{160mm}
\begin{tabular}{lrrrrlrrr}
\hline\noalign{\smallskip}
$\;\;$spectrum&$kT$&norm&$\Gamma$&$E_{\rm CO}$&norm&$E_{\rm l}$&norm&$\chi^2$/d.o.f.\\
&keV&&&keV\\
\noalign{\smallskip\hrule\smallskip}

Normal Branch\\
n1&2.05$\pm$0.08 &0.68$\pm$0.22 &1.89$\pm$0.28 &6.0$^{+1.3}_{-0.7}$  &181$^{+58}_{-34}$ &6.20$\pm$0.34&0.82$\pm$0.66 &41/54\\
n2&2.12$\pm$0.06 &0.80$\pm$0.26 &1.98$\pm$0.39 &6.4$^{+2.4}_{-0.9}$  &200$^{+92}_{-41}$ &6.11$\pm$0.35&1.03$\pm$0.60 &29/54\\
n3&2.16$\pm$0.05 &0.82$\pm$0.17 &1.90$\pm$0.23 &6.1$^{+0.7}_{-0.7}$  &191$^{+44}_{-31}$ &6.26$\pm$0.36&0.81$\pm$0.50 &30/54\\
n4&2.16$\pm$0.04 &0.92$\pm$0.15 &1.93$\pm$0.19 &6.4$^{+0.8}_{-0.6}$  &195$^{+57}_{-28}$ &6.28$\pm$0.33&0.74$\pm$0.40 &32/54\\
n5&2.21$\pm$0.08 &1.07$\pm$0.20 &2.00$\pm$0.27 &6.5$^{+1.0}_{-0.7}$  &215$^{+54}_{-36}$ &6.31$\pm$0.19&0.81$\pm$0.15 &34/54\\
\hline\noalign{\smallskip}
Flaring Branch\\
f1&2.02$\pm$0.07 &0.70$\pm$0.20 &1.85$\pm$0.27 &5.7$^{+1.0}_{-0.6}$  &181$^{+51}_{-33}$ &6.23$\pm$0.35&0.89$\pm$0.54 &46/54\\
f2&1.96$\pm$0.05 &1.01$\pm$0.18 &1.99$\pm$0.22 &6.1$^{+0.9}_{-0.7}$  &208$^{+41}_{-35}$ &6.19$\pm$0.17&0.99$\pm$0.20 &45/54\\
f3&2.00$\pm$0.04 &1.22$\pm$0.23 &1.96$\pm$0.32 &5.9$^{+1.4}_{-0.8}$  &211$^{+70}_{-44}$ &6.18$\pm$0.36&1.11$\pm$0.70 &30/54\\
f4&2.01$\pm$0.06 &1.56$\pm$0.23 &1.98$\pm$0.29 &6.0$^{+1.1}_{-0.8}$  &227$^{+65}_{-46}$ &6.28$\pm$0.32&1.16$\pm$0.67 &42/54\\
f5&2.04$\pm$0.04 &1.87$\pm$0.22 &2.08$\pm$0.28 &6.6$^{+1.4}_{-0.8}$  &254$^{+77}_{-50}$ &6.25$\pm$0.30&1.31$\pm$0.84 &53/54\\
f6&2.05$\pm$0.03 &2.08$\pm$0.21 &2.11$\pm$0.26 &7.4$^{+1.4}_{-1.0}$  &267$^{+62}_{-38}$ &6.39$\pm$0.26&1.11$\pm$0.44 &29/48\\
f7&2.12$\pm$0.03 &2.73$\pm$0.23 &2.32$\pm$0.24 &7.6$^{+2.4}_{-1.1}$  &358$^{+80}_{-67}$ &6.30$\pm$0.19&1.46$\pm$0.32 &28/45\\
f8&2.22$\pm$0.02 &3.44$\pm$0.20 &2.50$\pm$0.23 &8.6$^{+1.8}_{-1.2}$  &457$^{+106}_{-78}$ &6.28$\pm$0.16&1.74$\pm$0.28&31/45\\
f9&2.40$\pm$0.02 &4.19$\pm$0.19 &2.54$\pm$0.20&10.3$^{+2.2}_{-1.6}$  &496$^{+95}_{-82}$  &6.34$\pm$0.16&2.13$\pm$0.36&50/47\\
\noalign{\smallskip}\hline
\end{tabular}\\
The normalization of the blackbody is in units of $\rm {10^{39}}$ erg s$^{-1}$ for
a distance of 10 kpc, the normalization of the cut-off power law is in units of
photon cm$^{-2}$ s$^{-1}$ keV$^{-1}$ at 1 keV and the line normalization has units of photon cm$^{-2}$ s$^{-1}$.
\end{minipage}
\end{center}
\end{table*}
systematic error of 2\% was applied to each channel instead of
the more normal 1\%, to avoid obtaining unrealistic values of $\chi^2$.
HEXTE spectra were also extracted for each selection using the GTI filters
from the PCA in each case using the {\it RXTE}-specific Ftool {\sc hxtlcurv}.
Each raw data file from Cluster 1 provided a source and a background spectrum, 
and the individual spectra were then added.

\subsection{Spectral fitting results}

To establish the upper limit to the usable energy range in each instrument, a source
spectrum (source + background) was compared with the background spectrum.
This showed that in general PCA data could be used up to $\sim$22 keV 
(where the source flux was substantially greater than the background), and HEXTE
data could be used up to $\sim$50 keV. 
%Also, the response functions are not well-defined 
%below certain energies. 
The bands 2.5 -- 22 keV and 20--50 keV
were firstly applied in spectral fitting. However, inspection of individual spectra 
sometimes required further channels to be ignored so as to exclude data, for example, 
above 21 keV in the PCA if the Poisson errors became too large for sensible fitting.
An instrument response function was generated for the PCA 
using {\sc pcarsp} embodying the latest improvements to the response made in 2002 
by the release of Ftools 5.2. Use of the latest response function made a significant
difference to spectral fitting results. The appropriate HEXTE response files 
for Cluster 1 were applied (i.e. the {\it arf} file of May 26, 2000 and the {\it rmf} file
of March 20, 1997). 
%hexte\_00may26\_pwb013.arf and hexte\_97mar20c\_pwb013.rmf). 
The rmf file was rebinned to a total of 64 channels
to match the HEXTE data with the Ftool {\sc rbnrmf}.

PCA and HEXTE spectra were fitted simultaneously using {\it Xspec}
allowing a normalization factor for HEXTE with respect to the PCA. Parameter values were chained
between the PCA and HEXTE. Because the PCA does not extend
below 2.5 keV where the column density is largely defined, the Galactic column density
was set to the value $\rm {0.3\times 10^{22}}$ atom cm$^{-2}$ (Christian \& Swank 1997).
Simple one-component models were entirely unable to fit the spectra. Fitting a typical
spectrum (n1: on the Normal Branch close to the soft apex) gave $\chi^2$/d.o.f. = 
7099/60 for an absorbed power law, 6281/60 for an absorbed blackbody, 1340/60
for absorbed bremsstrahlung, and 98/59 for an absorbed cut-off power law, although
in the last case, the value of the normalization constant was unacceptable (close to
unity).

Next, the two-component model discussed in Sect. 1 was applied, consisting of an
absorbed blackbody and a cut-off power law: {\sc ab*(bb + cpl)}, the blackbody
originating on the neutron star, and the Comptonized emission in an extended accretion
disk corona. This model gave $\chi^2$/d.o.f. = 92/57 for spectrum n1; however, it
was clear from the residuals that a broad iron line was present, so the final
model tried was {\sc ab*(bb + cpl + gau)}, with a Gaussian line added. This model
gave $\chi^2$/d.o.f. =41/54, and provided excellent fits to all of the spectra.
The line half-width $\sigma$ was fixed at 0.8 keV, to avoid the tendency
of a broad line in spectral modelling to absorb the neighbouring continuum.
A typical fit (spectrum n1) is shown in Fig. 4, and the fitting results for all spectra shown in Table 1.
%The good fit shown in Fig. 4 shows that there is no a high energy tail in Sco\th X-1
%as claimed by d'Amico et al. (2001). These authors fitted a physically inappropriate
%bremmstrahlung model to Sco\th X-1 when it is known that Comptonization plays an
%important role in LMXB in general, and their detection of a high energy tail 
%is an artifact of this.

We also fitted a two-component model consisting of blackbody emission from the neutron star
plus the Zdziarski et al. (1996) Comptonization model (the {\sc thcompds} model in {\it Xspec}) 
with disk blackbody seed photons.
This model corresponds approximately to Comptonization in an extended ADC, except
that the seed photons are assumed to originate from the complete disk instead of
the inner 15\%. However, tests
have shown that this does not affect the spectrum above 0.1 keV. Spectral fitting results
using this model were essentially identical with those shown in Table. 1, with no discernible
differences in blackbody parameters for any spectrum, and only minor changes in power law index.
This test demonstrates that a cut-off power law representation of Comptonization
is valid given the extended nature of the ADC.

\begin{figure}[!h]                                            %Fig. 5
\includegraphics[width=68mm,height=88mm,angle=270]{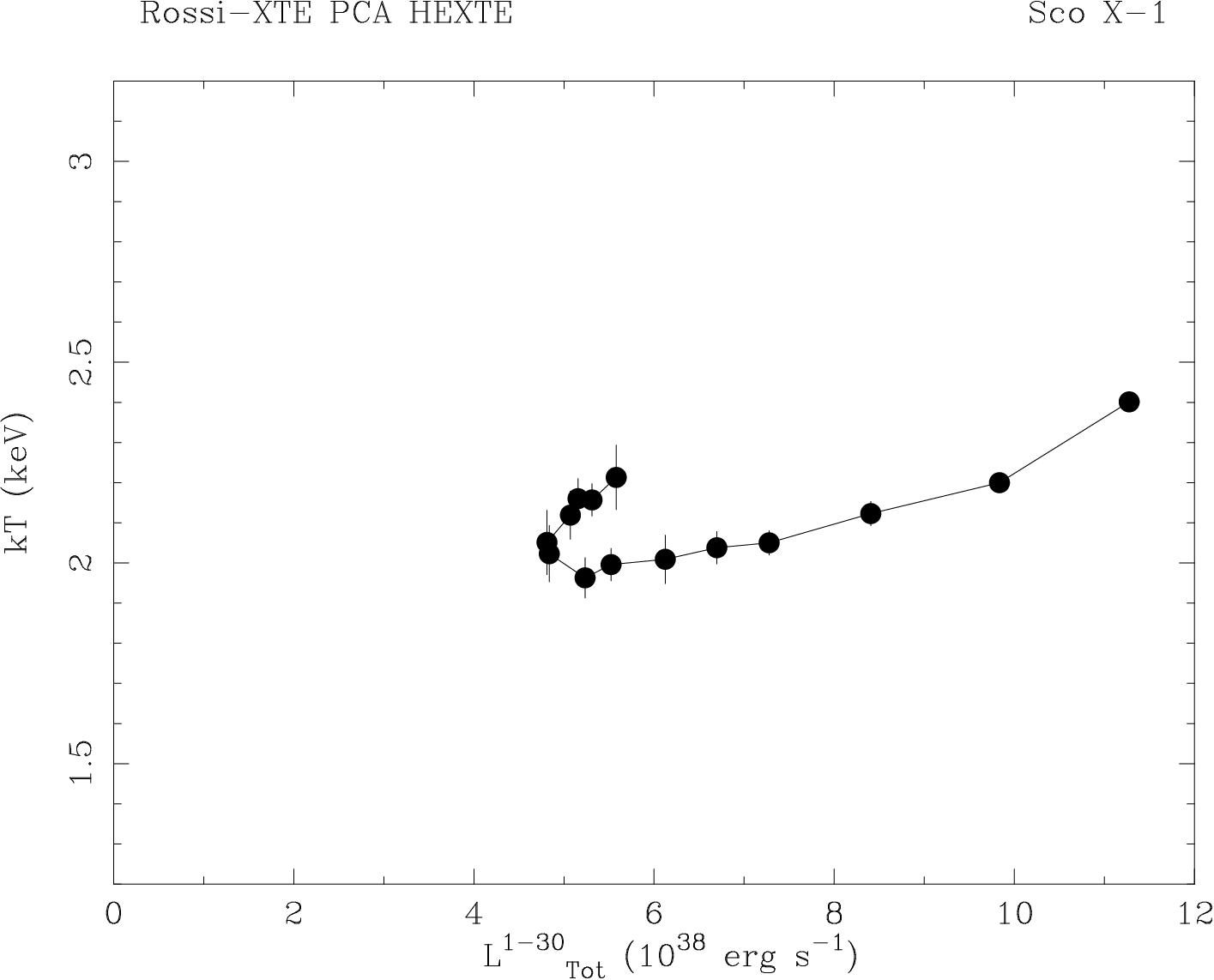}  % was kt
\includegraphics[width=68mm,height=88mm,angle=270]{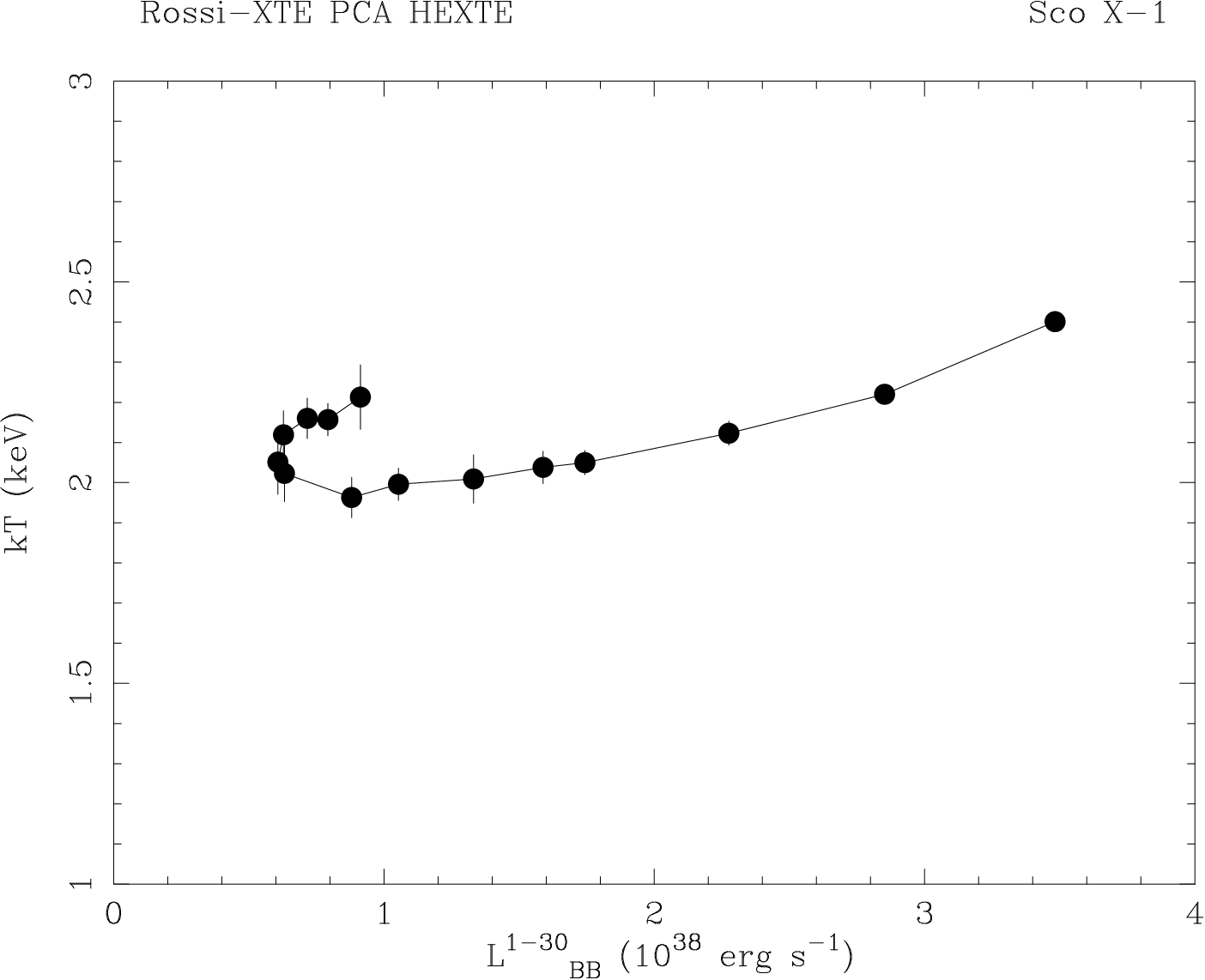} % was kt1
\includegraphics[width=68mm,height=88mm,angle=270]{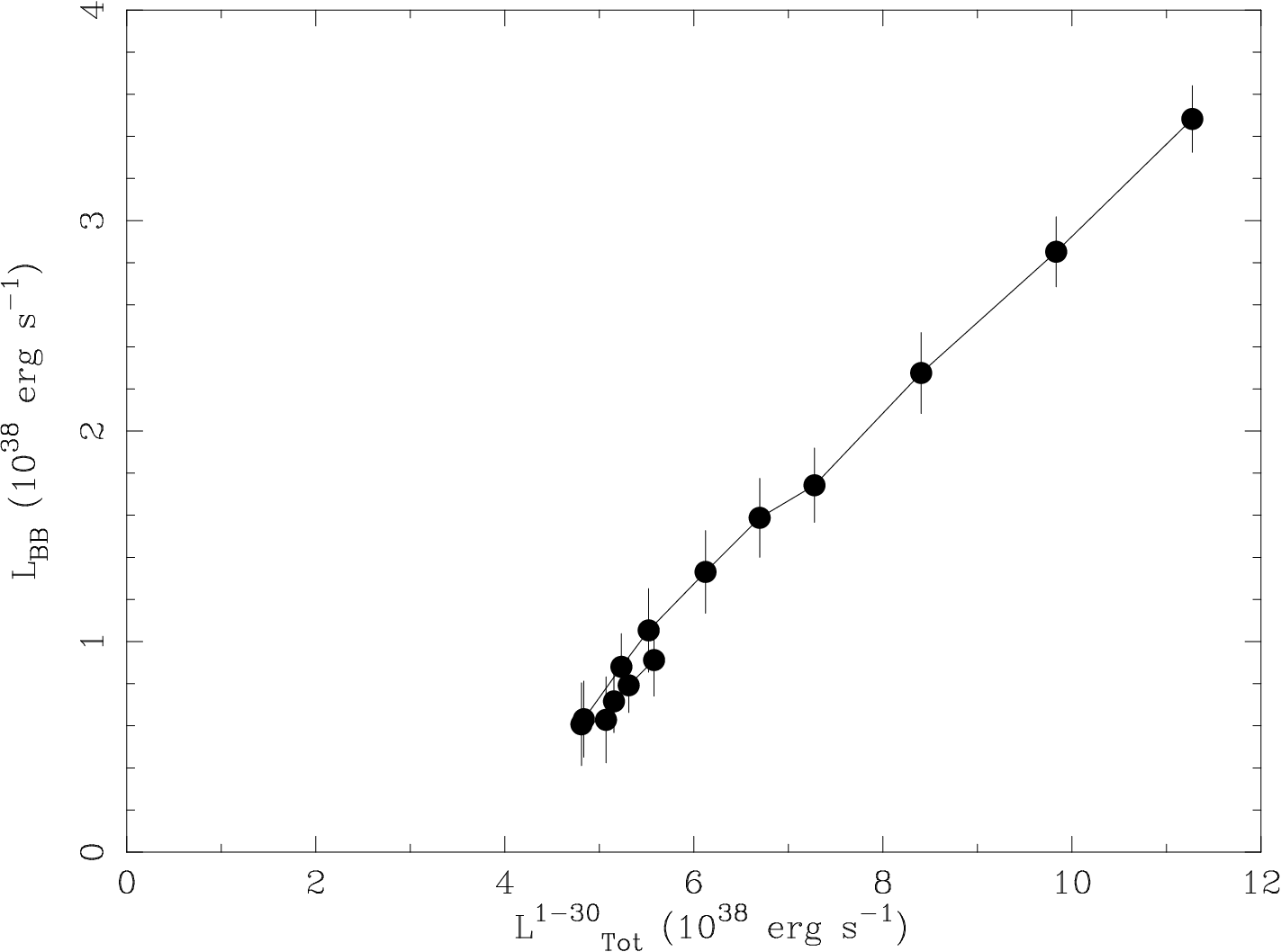} % was lbb
\caption{upper panel: blackbody temperature as a function of total luminosity; centre panel:
$kT$ as a function of blackbody luminosity; lower panel: blackbody luminosity {\it versus}
total luminosity. 90\% confidence uncertainties are shown}
\end{figure}

Spectral fitting results using the Birmingham model
for the blackbody emission component are shown in Fig. 5 as a
function of total luminosity in the band 1--30 keV, calculated assuming that the
distance of Sco\th X-1 is 2.8 kpc (Bradshaw et al. 1999). Firstly, the upper panel gives the
variation of blackbody temperature with total luminosity. It can be seen that as the source
moves along the Flaring Branch to higher intensities, $kT$ rises, rather slowly at first,
and then faster. In fact, during the first 40\% of the flaring movement, the increase in $kT$ 
is small so that the changes are close to being isothermal. However, the 
luminosity of the blackbody component $L_{\rm BB}$, obtained from the blackbody
flux in the band 1--30 keV, increases all along the Flaring Branch (Table 1), as can
be seen in  the lower panel. Thus, it is clear that the early stages of flaring are
dominated by an increase in blackbody area as the emitting region
spreads over the neutron star. This is accompanied by an increase of $kT$ to
2.4 keV at the peak of flaring. The central panel shows $kT$ as a function
of $L_{\rm BB}$, with a smooth progression of values along the tracks as
in the upper panel.

A smooth progression of $kT$ values also takes place in movement along the
Normal Branch towards the soft apex. It is immediately apparent 
from the upper panel that the total luminosity decreases as the source moves along 
the Normal Branch towards the apex. The lower panel shows that $L_{\rm BB}$ as well 
as $L_{\rm Tot}$ decreases during Normal Branch movement towards the apex, i.e.
the luminosity of the Comptonized emission $L_{\rm CPL}$ also decreases. These results suggest
that $\dot M$ may actually decrease on the Normal Branch, contrary to the
widely-held view that $\dot M$ increases continually (see Sect. 6).
Of course, it has been known for a long time that the X-ray intensity decreases
on the Normal Branch, and this clearly presents a problem for the hypothesis that
$\dot M$ continually increases. One possibility is that the $\dot M$ inferred from the
luminosity is not the total, but only part of it (van der Klis 2000).

Finally, the lower panel shows that the blackbody luminosity bears a remarkably
simple relation to the total luminosity, that is close to linear, but not passing
through the origin, so that at the soft apex, $L_{\rm BB}$ is about 10\%
of  $L_{\rm Tot}$, this fraction increasing to 30\% at the peak of flaring.

\begin{figure}[!h] %Fig. 6
\includegraphics[width=68mm,height=88mm,angle=270]{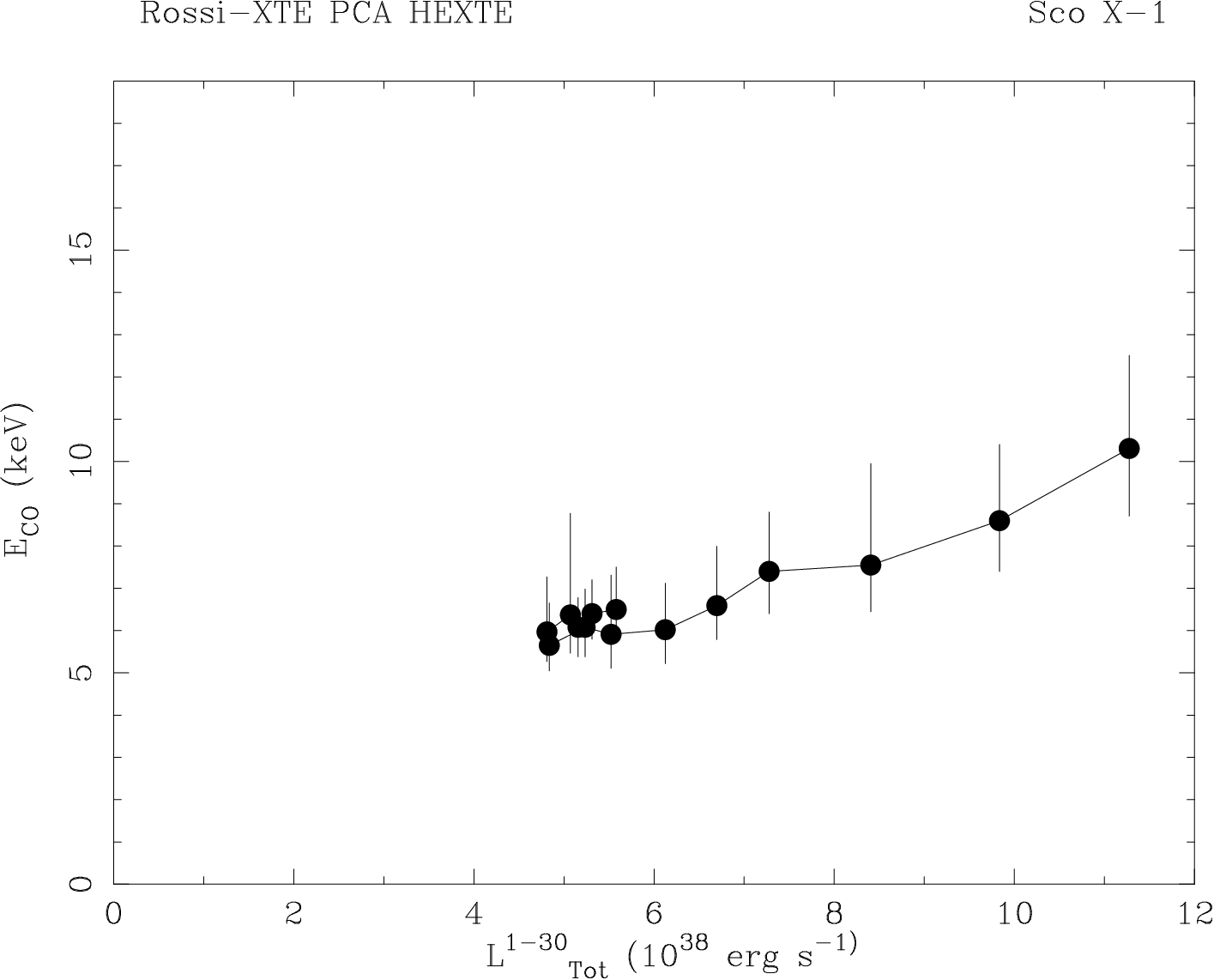} % was eco
\includegraphics[width=68mm,height=88mm,angle=270]{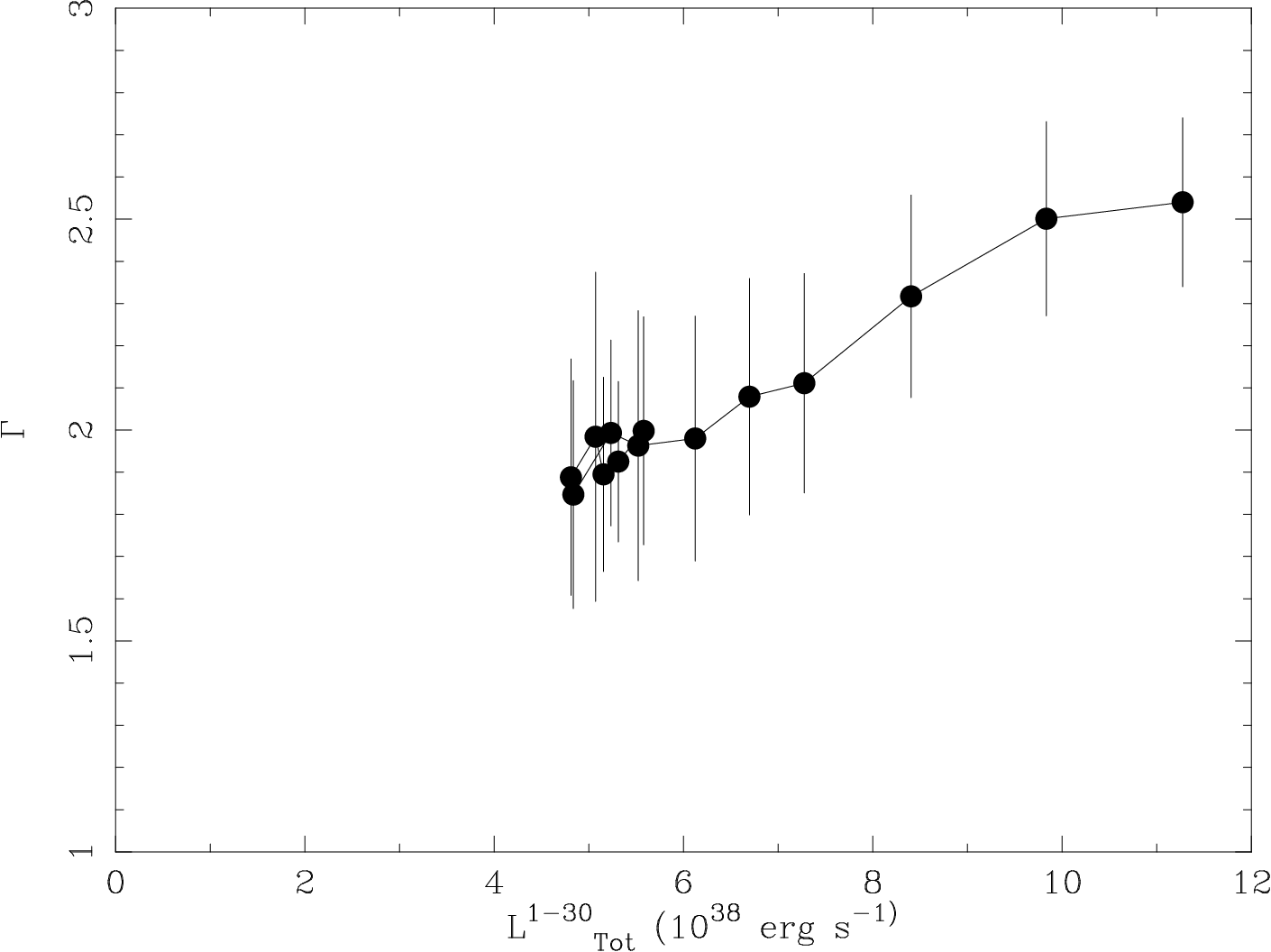} % was gamma
\includegraphics[width=68mm,height=88mm,angle=270]{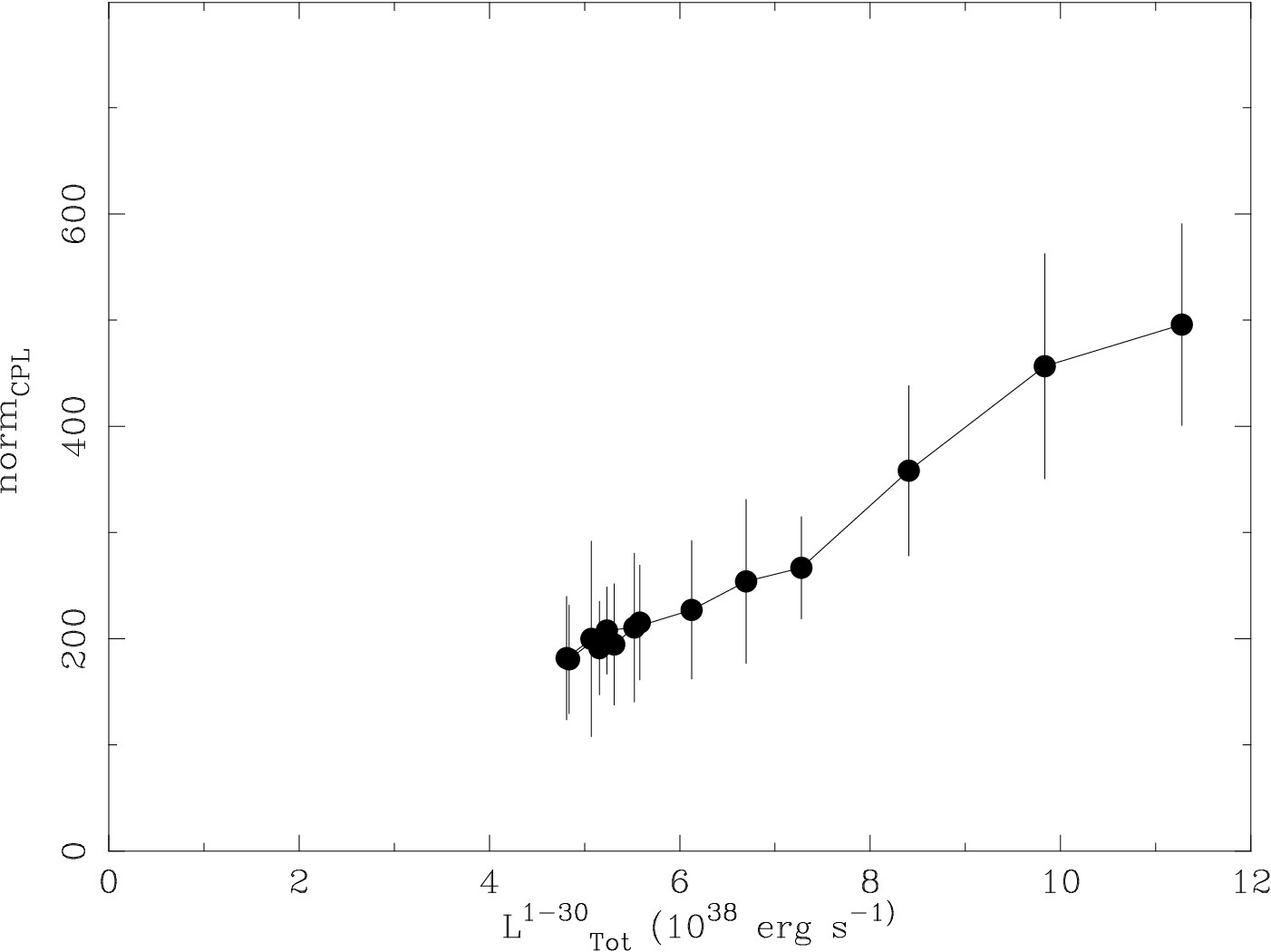} % was norm
\caption{upper panel: Comptonization cut-off energy; centre panel: power law photon index;
lower panel: normalization, all as a function of total luminosity. 90\% confidence
uncertainties are shown} 
\end{figure}

Properties of the Comptonized emission are shown in Fig. 6. The normalization (lower panel)
increases as flaring progresses. It is interesting that the Comptonization
cut-off energy $E_{\rm CO}$ also increases systematically in flaring, implying an
increase of electron temperature in the Comptonizing region, i.e. the extended ADC
above the accretion disk. This increase was not previously known in studies of
flaring.
There is also an increase in the power law photon index
$\Gamma $, from $\sim$1.8 at the soft apex to 2.5 at the peak of flaring; i.e. the
Comptonized emission becomes much softer. There is not strong evidence for
Normal Branch data being separate from Flaring Branch data, except possibly in the case
of the cut-off energy. The decrease of $E_{\rm CO}$ towards the soft apex
is supporting evidence that the blackbody luminosity does decrease on the
Normal Branch as the apex is approached, assuming that the value of $E_{\rm CO}$,
i.e. $T_{\rm e}$, does depend on $L_{\rm BB}$.

\begin{figure}[!h] %Fig. 7
\begin{center}
\includegraphics[width=68mm,height=88mm,angle=270]{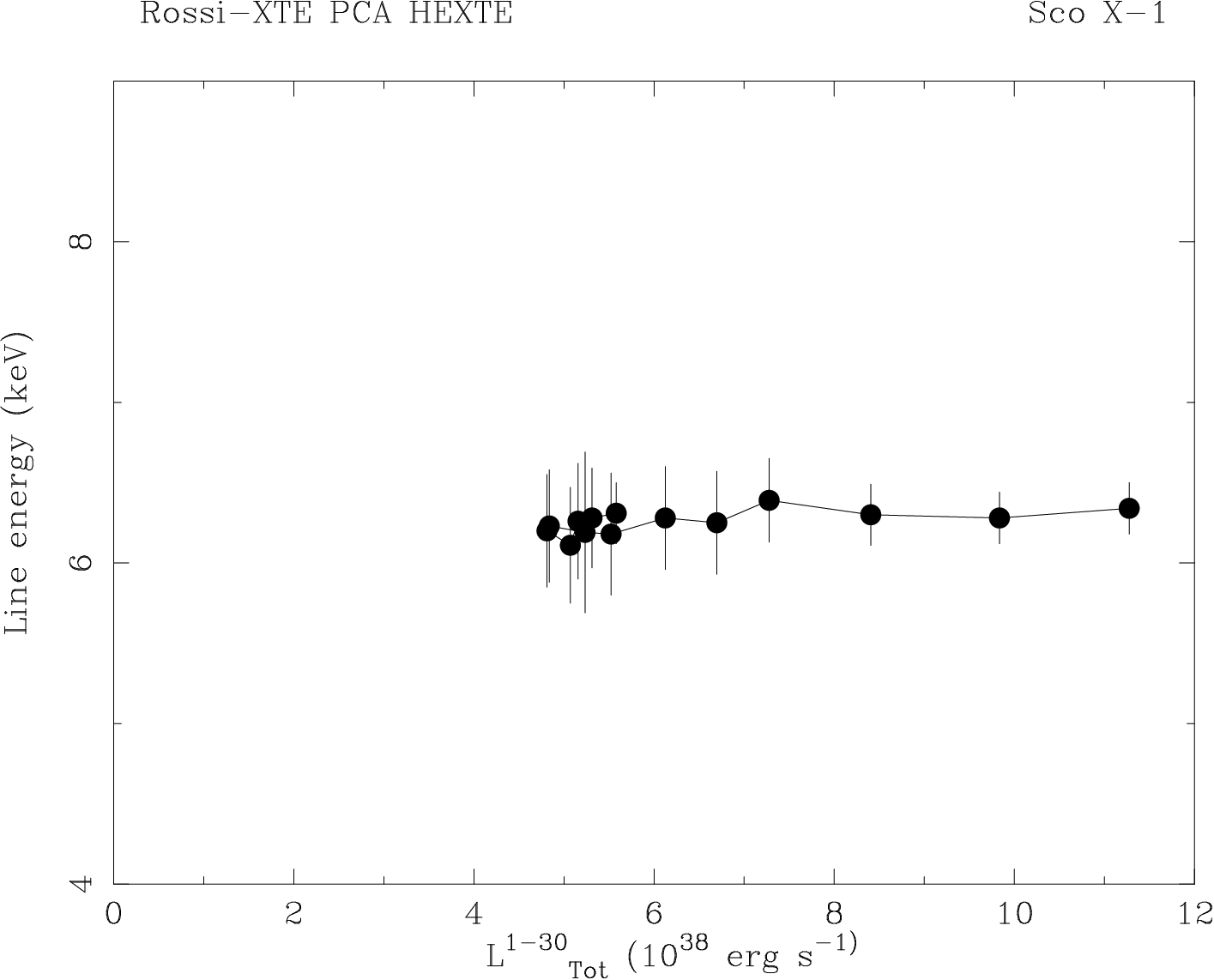} % was el
\includegraphics[width=68mm,height=88mm,angle=270]{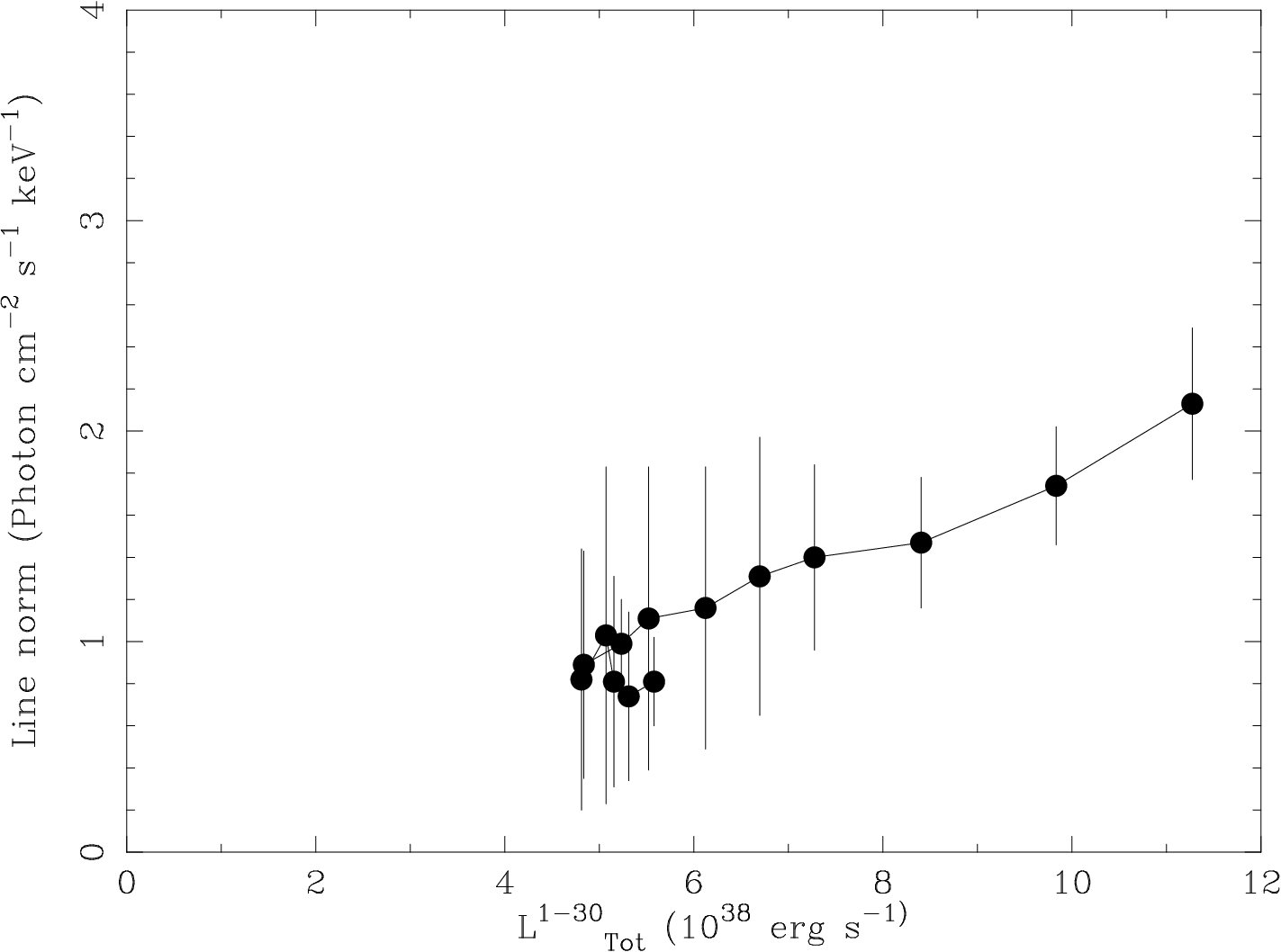} % was ln
\caption{Iron line energy (upper panel) and normalization
(lower panel)}
\end{center}
\end{figure}

In Fig. 7, we show the line energy and normalization; the line width $\sigma $
had been fixed in the fitting (Sect. 5.4). The line energy appears not to change
during Z-track movement; i.e. there is no evidence for change within the errors
and the energy is consistent with iron line emission at energy 6.4 keV, i.e.
fluorescent iron emission. The tendency of the line energy to be 
slightly low at $\sim$6.3 keV may suggest the possible presence of 
an absorption feature (Parmar et al. 2002).
The line normalization however, does change
significantly, increasing on the Flaring Branch. This suggests that
changes are due to the varying illumination by the neutron star blackbody.

\section{Discussion}

To discuss the changes taking place in the blackbody emission, we must first
summarize the evidence we presented in previous papers that the neutron star
blackbody emission is {\it not} substantially modified by electron scattering
and that results from using a simple blackbody model are thus valid.
In Sect. 1, we summarized the evidence that the blackbody emission in LMXB
originates on the neutron star, not from the inner accretion disk. 
The present results show a continuous pattern
of change in the blackbody $kT$ from the non-flaring to the flaring state,
and if we accept that X-ray flaring is analogous in high-luminosity sources
to X-ray bursting in fainter sources,
this is also evidence that the emission originates on the star.

In our study of X-ray flaring in the dipping LMXB X\th 1624-490 
(Ba\l uci\'nska-Church et al. 2001), we review the evidence for and against
possible modification of the neutron star blackbody. It has sometimes been claimed that
neutron star blackbody emission should be modified by electron scattering
in which case, fitting a simple blackbody could give incorrect results.
It was concluded that no substantial modification takes place, 
as the evidence for modification
is limited to a fraction of X-ray bursts in which $kT$ rises to apparently
super-Eddington values of 3 keV or more. In many bursts, and in non-burst emission
there is no evidence for modification. Moreover, whether modification takes
place depends strongly on the electron density in the atmosphere of the
neutron star which is poorly known. 

The present results show that flaring consists of strong changes in the blackbody
emission. We can
use the spectral fitting results to derive the blackbody emitting area
{\it via} Stefan's law: $L_{\rm BB}$ = 4$\pi$ $R_{\rm BB}^2\,\sigma\,T^4$.
In Fig. 8, we show the blackbody radius $R_{\rm BB}$ on which the emitting area depends.

\begin{figure}[!h] %Fig. 8
\begin{center}
\includegraphics[width=68mm,height=88mm,angle=270]{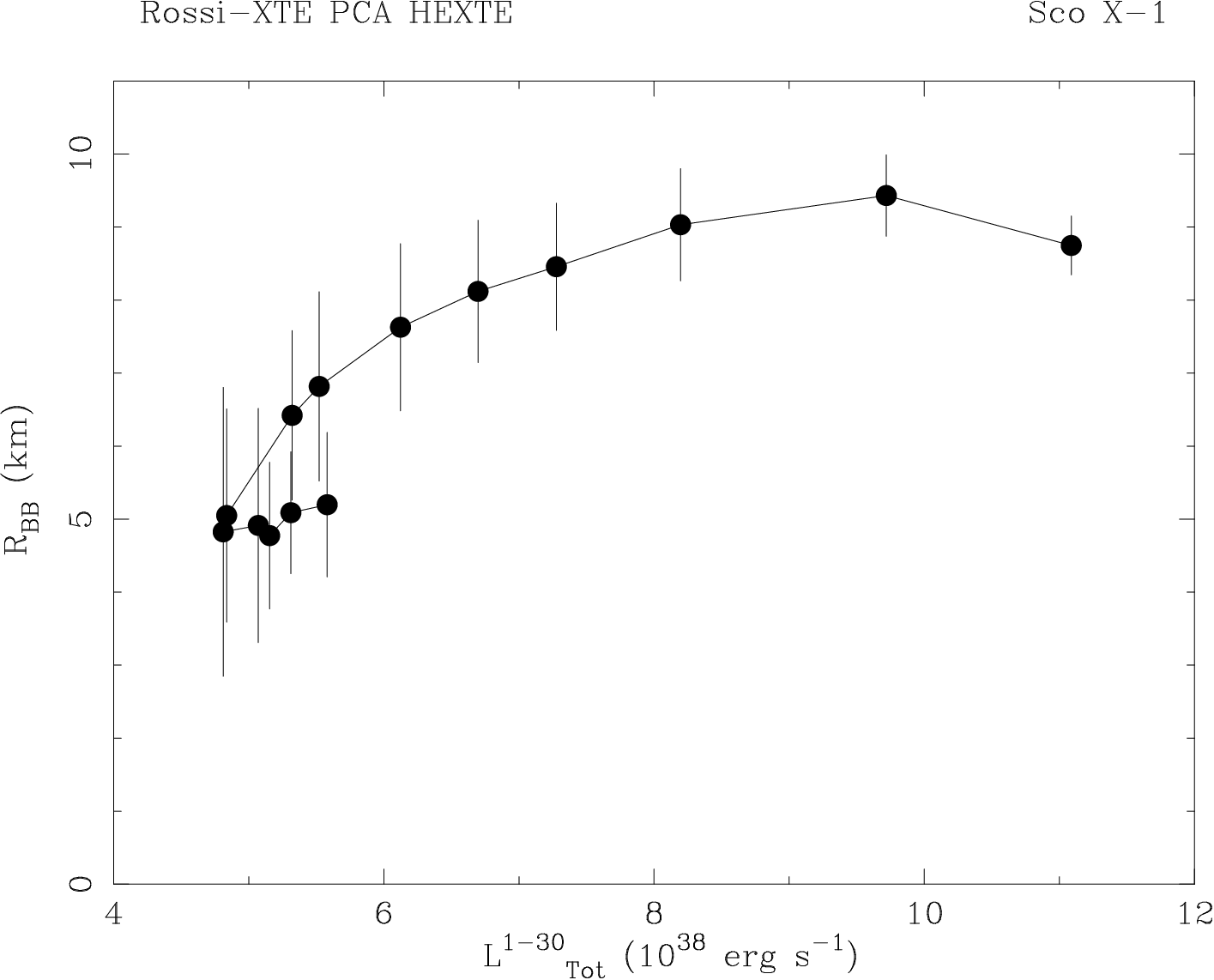} % was bb
\includegraphics[width=68mm,height=88mm,angle=270]{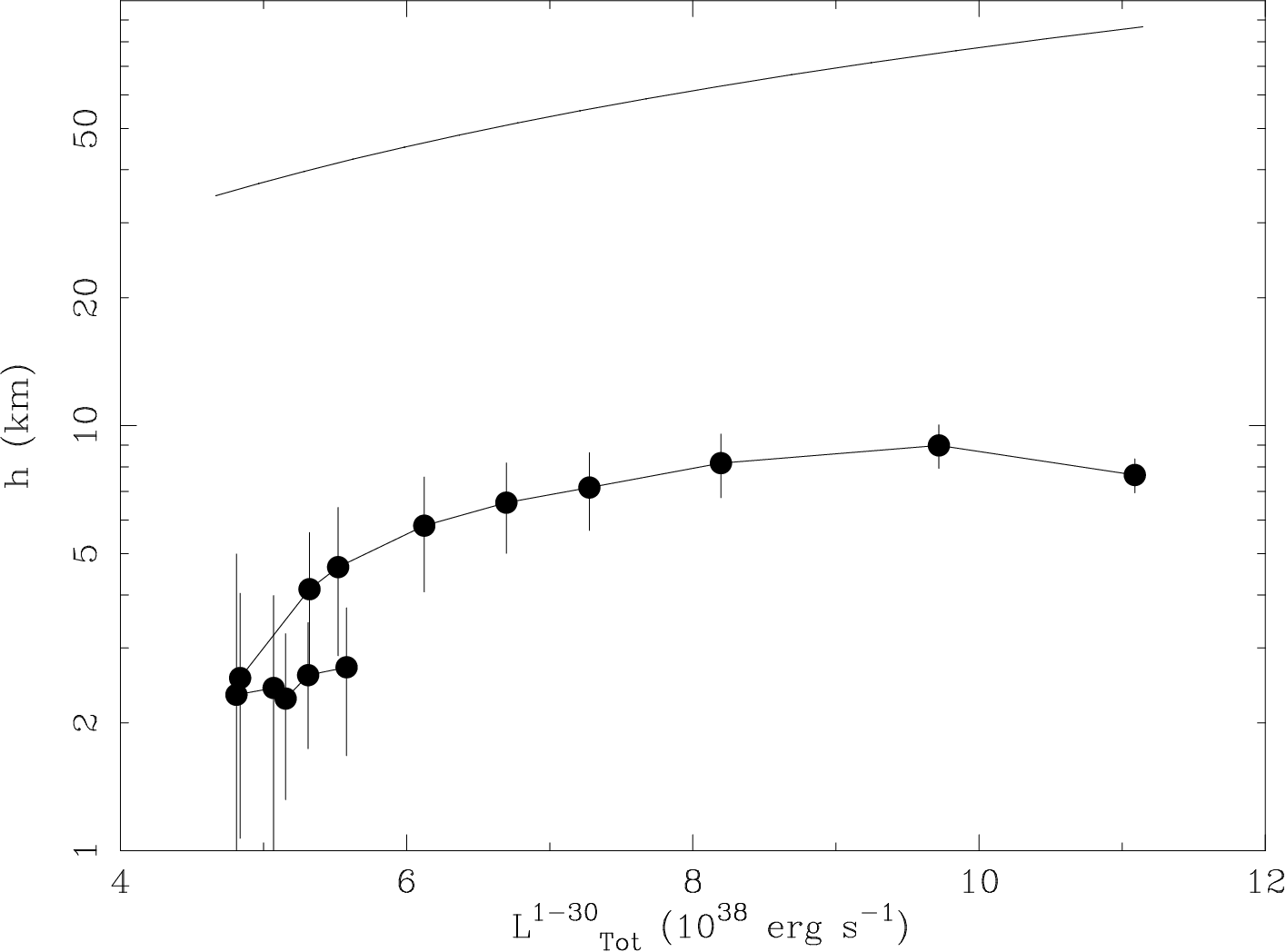} % was h
\caption{Upper panel: blackbody radius $R_{\rm BB}$; lower panel: half-height of the
neutron star emission region $h$ (see text). Also shown is the equilibrium height
of the radiatively-supported inner accretion disk (see text)}
\end{center}
\end{figure}

On the Flaring Branch, there is a substantial increase in blackbody radius up to
a maximum value, after which the radius decreases. The maximum value is 9.4$\pm$0.6 km,
consistent with the radius of the neutron star. Fig. 3 shows a break in the flaring
branch hardness-intensity diagram such that the flaring becomes harder after this
break ($\sim$ $\rm {2.3\times 10^{4}}$ count s$^{-1}$), and it is
clear that this break takes place when the emission covers the whole of the neutron 
star. The value of $kT$ increases at first slowly in flaring, then more rapidly.
%Since X-ray flaring takes place in more luminous sources not in general exhibiting
%X-ray bursting, we assume that nuclear energy is released on the surface of the
%star resulting in the initially restricted hot X-ray emitting region expanding
%to fill the star, after which the region becomes hotter at constant area.
While it has been known for many years that $kT$ increases in flaring
in sources similar to Sco\th X-1, the increase of $R_{\rm BB}$ 
to a maximum value of $\sim $10 km, has not been known. It may also be that
this does not always take place, and is a feature of the source when very bright.

On the Normal Branch, the blackbody radius decreases slightly as the soft apex 
is approached, while $kT$ falls, and the measured $L_{\rm Tot}$ decreases
substantially. The most interesting feature is the implied {\it decrease of
$\dot M$ on the Normal Branch}. This is contradictory to the view widely-held
that in motion along the Z-track from Horizontal to Flaring Branch,
$\dot M$ increases continuously. This was proposed by
Hasinger et al. (1989), who noted that this would however require $\dot M$
and intensity to be anti-correlated on the Normal Branch; i.e. it is a difficult
to explain increasing mass accretion rate but decreasing X-ray intensity.
Hertz et al. (1992) showed that the properties
of Sco\th X-1 depend on distance along the Z-track measured from the soft apex,
with positive values on the Flaring Branch and negative values 
on the Normal Branch. Thus, the smooth progression of properties implies
dependence on a single parameter, normally assumed to be $\dot M$.
Vrtilek et al. (1990) made ultraviolet line and continuum observations
of \hbox{Cyg\th X-2} simultaneously with {\it Ginga} X-ray observations, and
argued that the ultraviolet emission increased along the Z-track supporting 
the contention that $\dot M$ increases. However, the label of Flaring Branch 
had been assigned to data with strong variability due to X-ray dipping which 
thus may not have been real flaring branch data. Omitting these data
from the results would mean that there is no very dramatic correlation of UV
emission with track position. However, our results {\it do suggest}
that $\dot M$ decreases on the Normal Branch, since the luminosities of {\it both} 
spectral components decrease. Clearly, this possibility is important to 
the understanding of the Z-track sources.

We next compare these results for $R_{\rm BB}$ with results previously 
obtained: i) from the {\it ASCA} survey of LMXB (Church \& Ba\l uci\'nska-Church
2001), and ii) a study of X-ray flaring in the bright dipping LMXB X\th 1624-490
Ba\l uci\'nska-Church et al. 2001). In the {\it ASCA} survey, an empirical
relation was found between $h$, the half-height of the emission region on the
neutron star, and $H$ the equilibrium height height of the inner radiatively-
supported disk. Over more than 3 decades of luminosity covered by the survey sources, 
it was found that $h$ $\approx$ $H$, suggesting possible mechanisms that
determine the extent of the blackbody emitting area on the star. Comparison
of these results with the model of Inogamov \& Sunyaev (1999) was made by Church et al. 
(2002). However, at the time of the survey, it was clear that the Z-track sources
could not be represented by a single point on the $h$--$H$ diagram, and would
have to form a track.
In Fig. 8 (lower panel), we show the variation of $h$ with $L_{\rm Tot}$
for the present data on Sco\th X-1.
This is related to blackbody radius by 4$\pi$ $R_{\rm BB}^2$
= \hbox{4$\pi R\, h$}, where $R$ is the radius of the neutron star assumed to be 10 km,
so that $h$ = $R_{\rm BB}^2/R$. We also show on this diagram, the half-height of the
inner radiatively-supported accretion disk $H$, given by the expression 
$H$ = $3\sigma_{\rm T} \dot M/8\pi m_{\rm p} c$ (Frank et al. 1992),
where $\sigma _{\rm T}$ is the Thomson cross-section.
It can be seen that $H$ is substantially larger than the radius of the neutron star
for all data in Fig. 8, which would imply that all of the star should
be emitting. However, we would not expect an equality $h$ = $H$ for any of
the present data if we assume that $h$ cannot be larger than 10 km. It is possible
that closer equality between $h$ and $H$ took place on the Horizontal Branch
for which no data are available.
Alternatively, it may be that a simple calculation of the radiative disk
height is not valid at such high mass accretion rates, since the inner
disk may be modified by irradiation by the neutron star.

In the source, X\th 1624-490, strong flaring also takes place. This is a dipping
but not a Z-track source; however strong flaring has been found to take place in
this source (Ba\l uci\'nska-Church et al. 2001)
and has recently been discovered in other dipping LMXB
such as XB\th 1254-690 (Smale et al. 2002; Church et al. 2003). These sources
are known to have high inclination as dipping is observed, and are bright
such that flaring takes place. However, it is not yet clear
whether the physical changes taking place in flaring are the same as in 
Sco\th X-1 and other Z-track sources which do not have so high inclination.
In the case of X\th 1624-490, it was shown that in the non-flaring source,
there was an equality $h$ = $H$; however, as flaring progressed, the emitting area
on the neutron star was strongly reduced falling to 10 times smaller than $H$
(Ba\l uci\'nska-Church et al. 2001). It was suggested that the increase of
radiation pressure of the neutron star blackbody emission in flaring had a
strong effect on the inner disk, blowing away the top layers and reducing the
vertical extent of the disk. This would then be expected to feed back onto the
neutron star leading to a reduced emission region height $h$, explaining
the observed decrease. This effect took place at $L_{\rm Tot}$ $\approxgt$ 
$\rm {1.2\times 10^{38}}$ erg s$^{-1}$. In the case of Sco\th X-1, the smallest
luminosity is $\rm {4.5\times 10^{38}}$ erg s$^{-1}$, substantially larger
so it may be the case that radiation pressure effects have taken place before the 
source got to its starting position in the present data. If a depression
in emitting area did take place, it is not unreasonable to expect the area
to increase again in flaring.

% does Mdot increase
%An interesting question is that of whether $\dot M$ increases during
%X-ray flaring. The increase of $L_{\rm Tot}$ could be due to the nuclear
%energy released once the conditions of density and temperature required
%for nuclear burning are achieved, i.e. when sufficient matter has
%accumulated by accretion on the surface of the star. At this point,
%the increases in luminosity may be fuelled by nuclear burning at constant
%$\dot M$, and it is not easy to conclude whether this is the case or not.
%Evidence for changing $\dot M$ might be found in the other emission
%component, i.e. the Comptonized emission from the ADC.
At the soft apex, $L_{\rm Tot}$ = $\rm {4.5\times 10^{38}}$ erg s$^{-1}$
which is super-Eddington for a 1.4M$_{\sun}$ neutron star for which $L_{\rm Edd}$
$\sim $ $\rm {2\times 10^{38}}$ erg s$^{-1}$, but depends on opacity and
composition (Paczynski 1983).
The measured total luminosity at the peak of flaring at
$\rm {1.1\times 10^{39}}$ erg s$^{-1}$ is extremely super-Eddington; however,
$L_{\rm BB}$ reaches a level of $\sim 4\times 10^{38}$ erg s$^{-1}$ 
which is more consistent with the Eddington limit.
The decrease in blackbody radius after 
the peak of flaring may thus be related to radiation pressure effects.
Sco\th X-1 is one of a very small number ($\sim$4) of persistent Galactic low
mass neutron star binaries that are, at times, substantially super-Eddington
(Grimm et al. 2002; Christian \& Swank 1997; Church \& Ba\l uci\'nska-Church 2001).
The above results depend of course, on the distance of Sco\th X-1. 
The distance to Sco\th X-1 was not well known, but a distance of
1.5 kpc was consistent with a luminosity of $\sim$10$^{38}$ erg s$^{-1}$
(White et al. 1985). However, Bradshaw et al.
(1999) from parallax measurements of the radio emission of Sco\th X-1 
in a series of eight VLBA observations
spanning 3 years obtained a parallax of 0.00036$\pm$0.00004$\arcsec$,
and a distance of 2.8$\pm$0.3 kpc which we have adopted. This has the
consequence that the luminosity of
Sco\th X-1 can be several times the Eddington limit.

Finally, we review the changes in the Comptonized emission and iron
line emission during track movement. As seen previously, a striking feature
is the increase of Comptonization cut-off energy in flaring implying an
increase of electron temperature in the Comptonizing region by a factor of two.
This is the first time that such heating effects on the ADC plasma due to
flaring have been seen. There is also a systematic increase in power law
photon index to $\sim$2.5 at the peak of flaring. The line energy does
not change and at $\sim $6.4 keV implies fluorescent emission. This may
originate in the ADC itself, providing the density is high so that
the ionization parameter is sufficiently low (see Ba\l uci\'nska-Church et al. 2001) or
possibly in the disk below the ADC. The increase of the line normalization
in flaring could be explained either by the effects of illumination of the 
emission region by the neutron star or by a change of $\dot M$.
These aspects will be further investigated in future work.

\begin{acknowledgements}
This work was partially supported by \hbox{Polish} KBN grant PBZ-KBN-054/P03/2001
\end{acknowledgements}


\begin{thebibliography}{}

\bibitem[]{}
Ba\l uci\'nska-Church, M., Church, M. J., Oosterbroek, T., et al.
1999, A\&A, 349, 495
%1323sax

\bibitem[]{}
Ba\l uci\'nska-Church, M., Humphrey, P. J., Church, M. J., \& Parmar, A. N.
2000, A\&A, 360, 583
%1624sax

\bibitem[]{}
Ba\l uci\'nska-Church, M., Barnard, R., Church, M. J., \& Smale, A. P.
2001, A\&A, 378, 847
%1624xte flaring

\bibitem[]{}
Barnard, R., Ba\l uci\'nska-Church, M., Smale, A. P., \& Church, M. J.
2001, A\&A, 380, 494 

\bibitem[]{}
Bradshaw, C. F., Fomalont, E. B., \& Geldzahler, B. J. 1999,
ApJ, 512, 121

\bibitem[]{}
Bradt, H. V., Rothschild, R. E., \& Swank, J. H. 1993, A\&AS, 97, 335

\bibitem[]{}
Canizares, C. R., Clark, G. W., Li, F. K., et al. 1975, ApJ, 197, 457

\bibitem[]{}
Church, M. J. 2001, Proceedings of 33rd Scientific Assembly of COSPAR, Warsaw,
July 2000, Adv Space Res, 28, 323

\bibitem[1995]{}
Church, M. J., \& Ba\l uci\'nska-Church, M. 1995, A\&A, 300, 441
%1624exo

\bibitem[]{}
Church, M. J., \& Ba\l uci\'nska-Church, M. 2001, A\&A, 369, 915
% Atoll sources

\bibitem[1997]{}
Church, M. J., Dotani, T., Ba\l uci\'nska-Church, M., et al. 1997, ApJ, 491, 388
% 1916 asca

\bibitem[1998]{}
Church, M. J., Ba\l uci\'nska-Church, M., Dotani, T., \& Asai, K. 1998a, ApJ, 504, 516
%0748

\bibitem[1998]{}
Church, M. J., Parmar, A. N., Ba\l uci\'nska-Church, M., et al. 1998b, A\&A,
338, 556
% 1916 sax

\bibitem[1998]{}
Church, M. J., Inogamov, N. A., \& Ba\l uci\'nska-Church, M. 2002, A\&A, 390, 146

\bibitem[]{}
Church, M. J., Ba\l uci\'nska-Church, M., \& Smale, A. P. 2003, A\&A, in preparation

\bibitem[]{}
Christian, D. J., \& Swank, J. H. 1997, ApJS, 109, 177

\bibitem[]{}
Done, C., \.Zycki, P., \& Smith, D. A. 2002, MNRAS, 331, 453

%\bibitem[]{}
%D'Amico, F., Heindl, W. D., Rothschild, R. E., Gruber, D. E., 2001, ApJ 547, L147

\bibitem[]{}
Frank, J., King, A. R., \& Lasota, J.-P. 1987,
A\&A, 178, 137

\bibitem[]{}
Frank, J., King, A. R., \& Raine, D. 1992, ``Accretion Power in
Astrophysics'', Cambridge University Press
%H_eq formula

\bibitem[]{}
Grimm, H.-J., Gilfanov, M., Sunyaev, R. A. 2002, A\&A 391, 923
% list of super-Eddington Galactic XRB + Magellanic Clouds

\bibitem[]{}
Hasinger, G., Priehorsky, W. C., \& Middleditch, J. 1989,
ApJ, 337, 843
%1st use of "Z-track"; QPO vary on track

\bibitem[]{}
Hertz, P., Vaughan, B., Wood, K. S., et al. 1992, ApJ, 396, 201

\bibitem[]{}
Inogamov, N. A., \& Sunyaev, R. A. 1999, Astron Lett, 25, 269

\bibitem[]{}
Jahoda, K., Swank, J. H., Giles, A. B., et al.
Stark, M. J., Strohmayer, T., Zhang, W., Morgan, E. H., 
1996, SPIE, 2808, 59

\bibitem[]{}
Kuulkers, E., van der Klis, M., \& Vaughan, B. A. 1996, A\&A, 311, 197

\bibitem[]{}
Mason, K. O., Charles, P.A., White, N. E., et al.
%Culhane, J.L., Sanford, P.W., Strong, K. T., 
1976, BAAS, 8, 433
%Sco X-1 like sources

\bibitem[]{}
Mitsuda, K., Inoue, H., Nakamura, N., \& Tanaka, Y. 1989, PASJ, 41, 97
%Eastern model

\bibitem[]{}
Muno, M. P., Remillard, R. A., \& Chakrabarty, D. 2002, ApJ, 568, L35

\bibitem[]{}
Paczy\'nski, B. 1983, ApJ, 267, 315

\bibitem[]{}
Parmar, A. N., Oosterbroek, T., Boirin, L., \& Lumb, D. 2002,
A\&A, 386, 910
% low line energy implies abs line(s)

\bibitem[]{}
Psaltis, D., Lamb, F. K., \& Miller, G. S. 1995, ApJ, 454, L137

\bibitem[]{}
Smale, A. P., Church, M. J., \& Ba\l uci\'nska-Church, M. 2001, ApJ, 550, 962
% 1624 dip

\bibitem[]{}
Smale, A. P., Church, M. J., \& Ba\l uci\'nska-Church, M. 2002, ApJ, 581, 1286
% 1254 dip

\bibitem[]{}
van der Klis, M. 2000, Ann Rev Astron Astrophys, 38, 717
% inconsistency of I and Mdot increasing

\bibitem[]{}
van der Klis, M., Stella, L., White, N. E., Jansen, F., \& Parmar, A. N. 
1987, ApJ, 316, 411
%QPO vary along Z-track

\bibitem[]{}
van Paradijs, J., \& Lewin, W. H. G. 1986, Proceedings of the NATO Advanced Research Workshop,
Rottach-Egern, West Germany, June 17-20, 1985. Dordrecht, D. Reidel Publishing Co.
%bright/faint lmxb

\bibitem[]{}
Vrtilek, S. D., Raymond, J. C., Garcia, M. R., et al. 1990, A\&A, 235, 162

\bibitem[]{}
Walter, F. M., Mason, K. O., Clarke, J. T., et al.
%Halpern, J., Grindlay, J. E., Bowyer, S., Henry, J. P., 
1982, ApJ, 253, 67

\bibitem[1982]{}
White, N. E., \& Swank, J. H. 1982, ApJ 253, L61
%bulge

\bibitem[]{}
White, N. E., Mason, K. O., Sanford, P. W., Ilovaisky, S. A., \& Chevalier, C.
1976, MNRAS, 176, 91
%2 states in Sco X-1

\bibitem[]{}
White, N. E., Peacock, A., \& Taylor, B. G., 1985, ApJ, 296, 475

\bibitem[]{}
White, N. E., Peacock, A., Hasinger, G., et al.
%Mason, K. O., Manzo, G., Taylor, B. G., Branduardi-Raymont, G., 
1986, MNRAS, 218, 129
% 2-cpt modelling of Sco X-1 - earlier work selfComp Brems

\bibitem[]{}
White, N. E., Stella, L.,  \& Parmar, A. N. 1988, ApJ, 324, 363
%extra cpt in bright sources

\bibitem[]{}
Zdziarski, A. A., Johnson, W. N., \& Magdziarz, P. 1996, MNRAS, 283, 193

\end{thebibliography}
\end{document}